\documentclass[twocolumn,aps,prl,amsmath,floatfix]{revtex4}
\usepackage{graphicx}
\begin{document}
\title{
Non-Fermi-liquid phases in the two-band Hubbard model:\\
Finite-temperature exact diagonalization study of Hund's rule coupling}
\author{A.~Liebsch and T.~A.~Costi} 
\affiliation{Institut f\"ur Festk\"orperforschung, 
             Forschungszentrum J\"ulich, 
             52425 J\"ulich, Germany}
\begin{abstract}
The two-band Hubbard model involving subbands of different widths is 
investigated via finite-temperature exact diagonalization (ED) and 
dynamical mean field theory (DMFT). In contrast to the quantum Monte Carlo 
(QMC) method which at low temperatures includes only Ising-like exchange 
interactions to avoid sign problems, ED permits a treatment of Hund's 
exchange and other onsite Coulomb interactions on the same footing.
The role of finite-size effects caused by the limited number of bath 
levels in this scheme is studied by analyzing the low-frequency behavior 
of the subband self-energies as a function of temperature, and by 
comparing with numerical renormalization group (NRG) results for an
effective one-band model. 
For half-filled, non-hybridizing bands, the metallic and insulating 
phases are separated by an intermediate mixed phase with an insulating  
narrow and a bad-metallic wide subband. The wide band in this phase 
exhibits different degrees of non-Fermi-liquid behavior, depending on 
the treatment of exchange interactions. Whereas for complete Hund's 
coupling, infinite lifetime is found at the Fermi level, in the 
absence of spin-flip and pair-exchange, this lifetime becomes finite. 
Excellent agreement is obtained both with new NRG and previous 
QMC/DMFT calculations. These results suggest that-finite temperature 
ED/DMFT might be a useful scheme for realistic multi-band materials.
electron systems 
\end{abstract}
\maketitle

\section{1 \ Introduction}

Strongly correlated materials exhibit a wealth of fascinating 
physical phenomena associated with complex single-electron and 
many-electron interactions. Transition metal oxides, for example,
tend to have partially filled shells of highly correlated $d$ 
electrons, surrounded by complicated lattice geometries, with 
many atoms and electrons per unit cell. The theoretical description 
of the electronic properties of these materials is a challenging 
topic in condensed matter physics. Significant advances were 
achieved during recent years via the dynamical mean field theory 
(DMFT) \cite{rmp,reviews} which provides a treatment of 
single-electron and many-electron interactions on the same footing. 

For realistic materials, DMFT has been used extensively in 
combination with the quantum Monte Carlo (QMC) method \cite{qmc}. 
This approach has the advantage that it can be readily applied to 
systems consisting of two or more subbands. It has the drawback,
however, that, to avoid sign problems at low temperatures, it 
includes only Ising-like exchange interactions \cite{held}.
Improvements of the QMC method to include the full Hund's coupling 
are so far limited to $T=0$ or rather high temperatures 
\cite{arita,koga3,rubtsov}.  
In view of this limitation there is clearly a need to explore 
alternative methods that are applicable to multi-band materials 
and complete onsite exchange interactions.

The aim of the present work is two-fold: First, exact diagonalization 
(ED) \cite{ed,rmp} is proposed as a potentially highly useful impurity 
solver for finite-$T$ DMFT studies of realistic systems. The attractive 
feature of this approach is that, in contrast to finite-$T$ QMC, onsite 
Coulomb and exchange interactions are treated on the same basis. 
Second, we apply finite-$T$ ED/DMFT to a highly nontrivial system which 
has recently received considerable attention, namely, the two-band 
Hubbard model consisting of subbands of different widths
\cite{anisimov,epl,prl,koga1,prb70,koga2,koga3,prl05,inaba,knecht,
biermann05,ferrero,medici,arita,song}. The phase diagram of this 
system was recently evaluated in Refs.~\cite{prl05,inaba}. Here we focus 
on the electronic properties of the so-called orbital-selective phase
in which the narrow band is insulating while the wide band is still 
metallic. To examine the influence of finite-size effects in the ED
approach, and to investigate the electronic properties in the limit 
of low temperatures, we have also performed numerical renormalization 
group (NRG) DMFT calculations for a simplified two-band model which 
is particularly suited for the intermediate phase.

The main result of this paper is that the two-band ED/DMFT calculations 
provide a correct picture of the electronic properties of the Hubbard 
model involving nonequivalent subbands, including the unusual 
non-Fermi-liquid properties of the orbital-selective phase.  
This outcome is remarkable since, for computational reasons, the number 
of bath levels per impurity orbital is necessarily smaller than in 
analogous one-band models. Nevertheless, despite this limitation, 
the ED results are in good qualitative or, in some cases, quantitative 
agreement with the NRG results. In the Ising case they also agree very 
well with previous QMC/DMFT results \cite{prb70,biermann05}. 

The overall consistency with the NRG and QMC calculations suggests 
that finite-$T$ ED/DMFT might indeed be very useful in the future 
to analyze realistic materials, especially, if full diagonalization
is replaced by finite-$T$ Lanczos methods \cite{jaklic,capone}
in order to be able to deal with larger cluster sizes.

Hubbard models involving nonequivalent subbands are relevant for 
compounds such as Ca$_{2-x}$Sr$_x$RuO$_4$ and Na$_x$CoO$_2$, where, 
as a result of the nearly two-dimensional lattice structure, coexisting 
narrow and wide bands arise naturally. Thus, onsite Coulomb energies 
can be simultaneously large and small relative to the widths of 
important subbands. As a function of doping concentration, both 
materials give rise to a remarkably rich sequence of phases, including 
superconductivity and Mott insulating behavior \cite{maeno,nacoo}. 
Evidently, the competition between multiple kinetic energy 
scales and local Coulomb and exchange energies is an important 
feature of these strongly correlated systems.

A consistent treatment of Coulomb and exchange interactions in materials
of this kind is important since it has recently become clear that the 
Hund's coupling has a decisive influence on the nature of the Mott 
transition \cite{koga2,prl05,ferrero,medici,arita,pruschke}. 
In fact, the different treatments of exchange terms in earlier finite-$T$ 
QMC \cite{epl,prl,prb70} and zero-$T$ ED \cite{koga1} calculations have 
given rise to some confusion, with apparently contradictory results. 
As was clarified in Ref.~\cite{koga2} for $T=0$ and in Ref.~\cite{prl05} 
for $T>0$, however, the QMC and ED results are in agreement provided 
that exchange interactions are treated in the same manner. Thus, for 
full Hund's coupling the two-band Hubbard model exhibits successive 
first-order transitions. In striking contrast, in the absence of 
spin-flip and pair-exchange only the lower transition remains 
first-order \cite{prb70,prl05,knecht}.

Moreover, as will be discussed in detail below, the nature of the 
intermediate phase depends in a subtle manner on the treatment of 
exchange interactions. In this regard the ED and NRG results yield 
the following picture: For full Hund's coupling, the wide band has 
infinite lifetime at $E_F$ but does not satisfy Fermi-liquid criteria 
at finite frequencies. This finding is consistent with recent results 
obtained by Biermann {\it et al.} \cite{biermann05} in $T=0$ two-band 
ED/DMFT calculations. For Ising exchange, instead, correlations 
are significantly enhanced and the lifetime becomes finite even at 
$E_F$, in agreement with QMC calculations \cite{prb70,biermann05}.   

{\it
Thus, the Mott transition in the Hubbard model involving different
subbands does not consist of equivalent sequential transitions. Instead,
when the narrow band becomes insulating, the wide band is forced into a
bad-metallic state whose deviations from Fermi-liquid behavior depend
in a qualitative manner on the treatment of exchange interactions.} 
         
In the past, finite-$T$ ED/DMFT methods have been applied mainly 
to single-band cases, where the incorporation of an appropriate 
cluster of bath levels is feasible. Throughout this paper we consider 
two impurity levels, each surrounded by either 2 or 3 bath levels, 
giving total cluster sizes $n_s=6$ or $n_s=8$ per spin. Below we 
demonstrate that even a cluster size of $n_s=6$ provides qualitatively 
correct subband self-energies. For instance, at $T=10$~meV the critical 
Coulomb energies of the orbital-selective Mott transitions for 
$n_s=6$ differ by only about {0.1\ldots0.2}~eV from those derived 
for $n_s=8$. This finding is interesting since it suggests that 
finite-$T$ ED/DMFT calculations for more realistic three-band models 
using a cluster size of $n_s=9$ (three impurity levels, each coupled
to two bath levels) should be useful. This would allow one to 
re-examine the Mott transition in systems that have been studied 
previously using QMC/DMFT for Ising exchange interactions.

A more accurate representation of low-frequency properties at low
temperatures requires three bath levels per impurity orbital: one 
near the Fermi level to provide adequate metallicity, and two for 
the upper and lower Hubbard bands, giving $n_s=8$ per spin. This 
extension leads to a significant reduction of finite-size effects.
Several comparisons of results for $n_s=6$ and $n_s=8$ are provided 
below to illustrate the range of applicability of the two-band 
ED/DMFT approach.    

The outline of the paper is as follows: In Section 2 the multiband
Hubbard model is specified and its numerical solution via the 
finite-$T$ exact diagonalization method is discussed. Section 3
describes the effective one-band model used in the NRG approach
to evaluate the electronic properties of the wide band in the 
intermediate phase when the narrow band is insulating. 
In Section 4 the ED/DMFT is applied to the purely metallic phase
just below the first Mott transition. 
Section 5 deals with the intermediate phase in the presence of full
Hund's coupling, where wide band exhibits infinite lifetime at
$E_F$, combined with non-Fermi-liquid behavior at finite frequencies.
In Section 6 this intermediate phase is considered in the absence of
spin-flip and pair-exchange terms, giving rise to finite lifetime 
even at $E_F$. In Section 7 the two-band ED approach is applied 
to the case studied previously within QMC/DMFT \cite{prb70}.
Section 8 presents analogous results for the case considered 
recently within QMC/DMFT by Biermann {\it et al.} \cite{biermann05}. 
Section 9 contains a brief discussion of iterated perturbation 
theory (IPT) DMFT results for the two-band model and Section 10 
provides a summary and outlook.             

\section{2 \ Multiband exact diagonalization method} 

The two-band Hubbard model for non-equivalent subbands is represented
by the Hamiltonian:
\begin{eqnarray}
   H &=&  H_0 \ + \ H_1 \ + \ H_2 \             \nonumber\\ 
 H_0 &=& \sum_{lmi\sigma}  t_{lmi} 
             c_{li\sigma}^+ c_{mi\sigma}  \nonumber\\
 H_1 &=& \sum_{li} U n_{li\uparrow} n_{li\downarrow} \ + \
         \sum_{l\sigma\sigma'} (U'-J\delta_{\sigma\sigma'}) 
                 n_{l1\sigma} n_{l2\sigma'}                \nonumber
\end{eqnarray}
\begin{eqnarray}
 H_2 &=& -J'\sum_l [c_{l1\uparrow}^+ c_{l1\downarrow}
            c_{l2\downarrow}^+ c_{l2\uparrow}    +  {\rm H.c.} ] 
                                                            \nonumber\\
     &&  -J'  \sum_l [c_{l1\uparrow}^+ c_{l1\downarrow}^+                   
                  c_{l2\uparrow} c_{l2\downarrow}    + {\rm H.c.} ]\, ,
\end{eqnarray}
where $c_{li\sigma}^+$ and $c_{li\sigma}$ are creation and
annihilation operators for electrons at site $l$ in orbital $i=1,2$
with spin $\sigma$ and  $n_{li\sigma}=c_{li\sigma}^+ c_{li\sigma}$. 
H.c.~denotes Hermitian conjugate terms.
$H_0$ is the single-particle Hamiltonian which we represent by 
half-filled, non-hybridizing bands with density of states 
$N_i(\omega)= 2/(\pi D_i)\sqrt{1- (\omega/D_i)^2}$ and widths 
$W_1=2$~eV, $W_2=4$~eV, where  $W_i=2D_i$. $H_1$ represents the 
anisotropic, Ising-like onsite Coulomb and exchange interactions 
and $H_2$ additional spin-flip and pair-exchange contributions. For 
isotropic Hund's coupling one has $J'=J=(U-U')/2$. Below, we discuss 
results for $J'=J$ as well as $J'=0$. The latter case corresponds to 
the Ising-like exchange treatment in previous QMC/DMFT calculations 
\cite{prl,prb70}.

The ED/DMFT results are derived from a two-band generalization of the 
approach employed for single bands \cite{ed,rmp}. 
The effective Anderson impurity Hamiltonian includes impurity levels 
$\epsilon_{1,2}$ and bath levels $\epsilon_{k=3\ldots n_s}$ 
(see Fig.~1). For particle-hole symmetry $\epsilon_{1,2}=0$. 
Each impurity level interacts with its own bath, so that for $n_s=6$  
impurity level 1 couples to bath levels 3,4 with $\epsilon_4=-\epsilon_3$, 
while level 2 couples to bath levels 5,6 with $\epsilon_6=-\epsilon_5$. 
For $n_s=8$ impurity levels 1,2 couple in addition to the bath levels 
$\epsilon_7=0$ and $\epsilon_8=0$, respectively.

\begin{figure}[b!]
  \begin{center}
  \includegraphics[width=6.0cm,height=3cm,angle=-0]{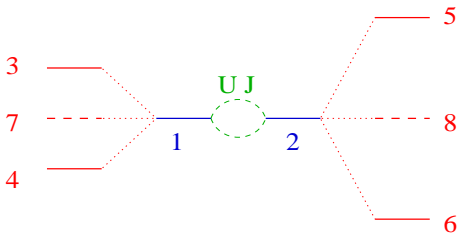}
  \end{center}
  \vskip-2mm
\caption{\label{ED}
Level diagram for ED scheme. Blue lines: impurity levels;
red solid (plus dashed) lines: bath levels for $n_s=6$ ($n_s=8$). 
Green lines: Coulomb and exchange interactions between impurity
levels; red dotted lines: hopping interactions between impurity
and bath levels.
}\end{figure}

Since at $T>0$ all states of the impurity Hamiltonian are used in the 
construction of the subband Green's functions, for $n_s=6$ the largest 
matrix to be diagonalized corresponds to the sector $n_\uparrow=
n_\downarrow=3$, with dimension $[n_s!/((n_s/2)!)^2]^2=400$. 
For $n_s=8$ this dimension increases to $4900$. 
We neglect hybridization between bands, so that self-energies and Green's 
functions are diagonal in orbital space. Denoting eigenvalues and 
eigenvectors of the impurity Hamiltonian by $E_\nu$ and $|\nu \rangle$, 
the 
subband Green's functions are evaluated from the expression 
\begin{equation}
 G_i(i\omega_n) = \frac{1}{Z} \sum_{\nu\mu} 
            \frac{\vert\langle \nu|c_{0i\sigma}^+ |\mu \rangle \vert^2}
                    {E_\nu - E_\mu - i\omega_n}
   [e^{-\beta E_\nu} + e^{-\beta E_\mu}].    
\end{equation}
Here, $\beta=1/k_B T$, $\omega_n=(2n+1)\pi/\beta$ are Matsubara 
frequencies, and $Z=\sum_\nu {\rm exp}(-\beta E_\nu)$ is the partition 
function. $l=0$ denotes the impurity site. Since we consider only 
paramagnetic phases the spin index of Green's functions and 
self-energies is omitted for convenience. Because of the particle-hole
symmetry of the present system, Green's functions and self-energies
are purely imaginary quantities. 

\begin{figure}[t!]
  \begin{center}
  \includegraphics[width=5.0cm,height=8cm,angle=-90]{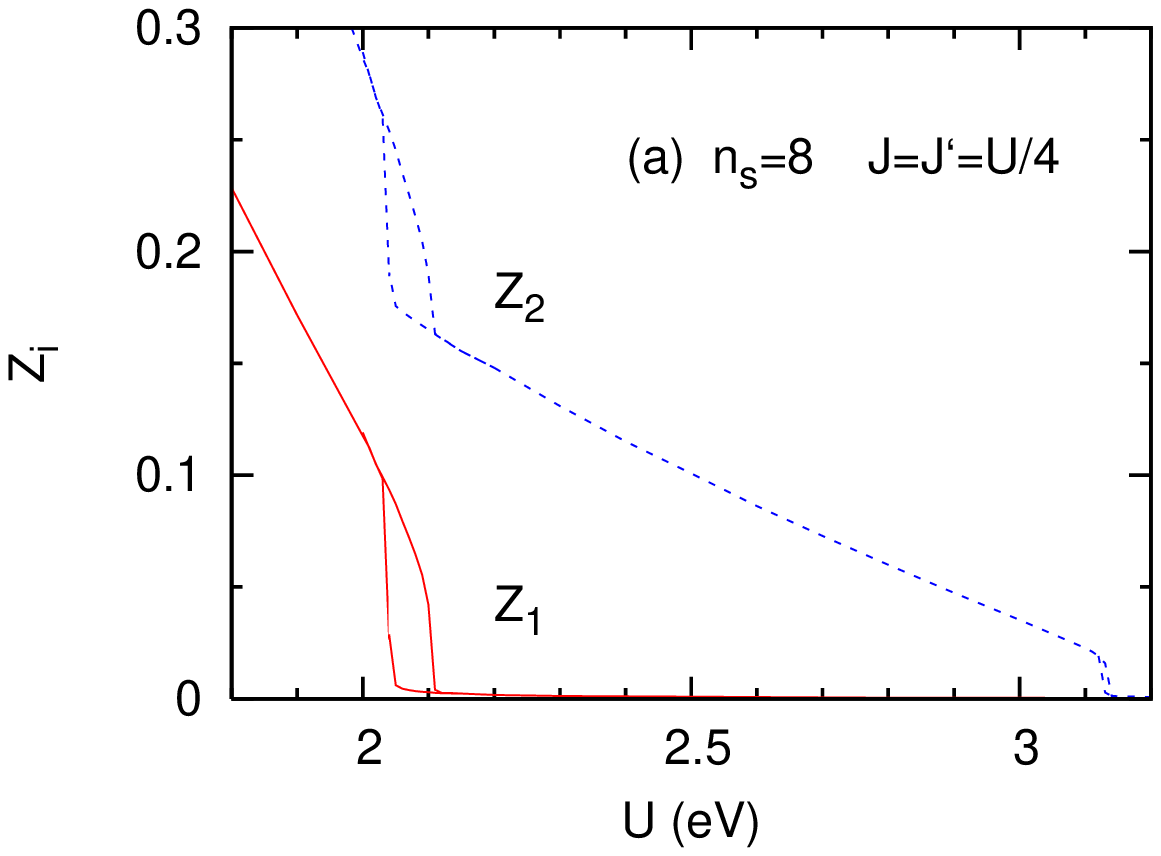}
  \includegraphics[width=5.0cm,height=8cm,angle=-90]{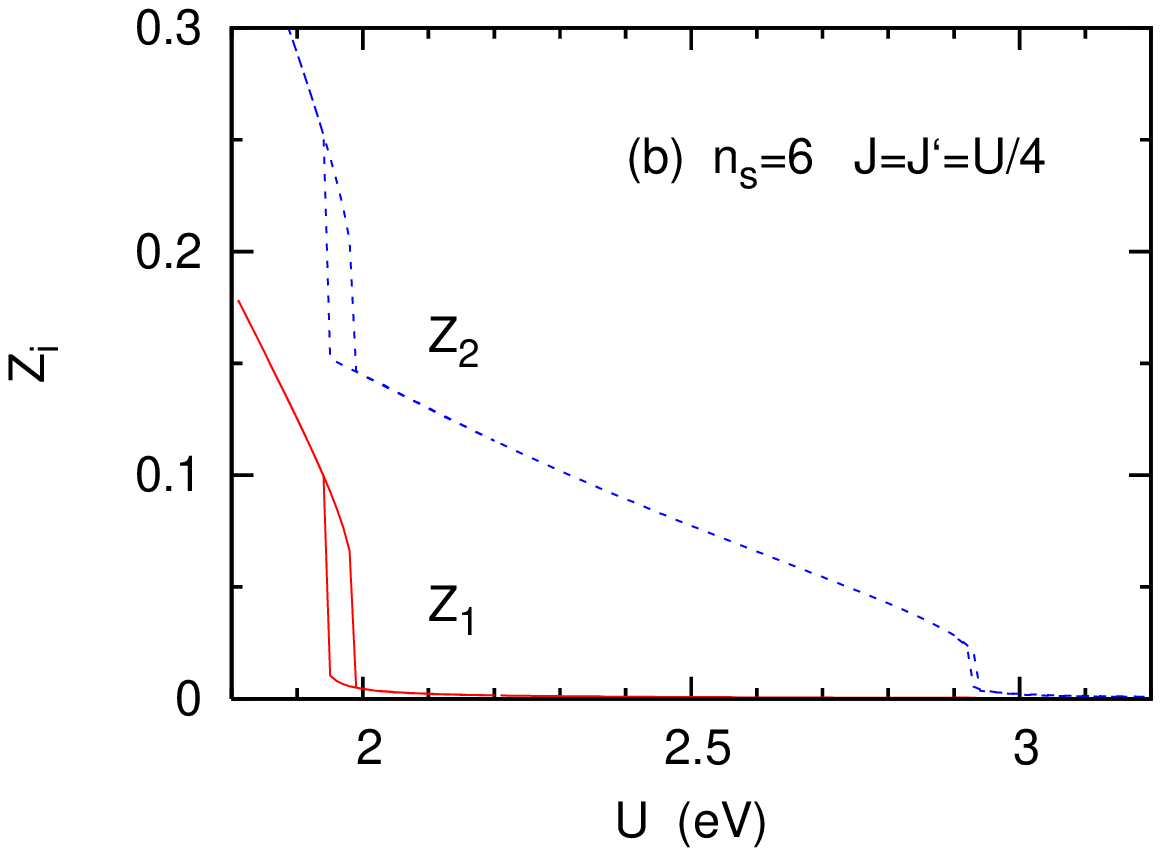}
  \end{center}
  \vskip-2mm
\caption{\label{ZU}
$Z_i(U)$ as a function of $U$ for $J=J'=U/4$ at $T=10$~meV, 
calculated within ED/DMFT for different cluster sizes:
(a) $n_s=8$; (b) $n_s=6$. 
Solid (red) curves: narrow band, dashed (blue) curves: wide band.  
}\end{figure}

To provide an overview of the Mott transitions in this model and 
to illustrate the sensitivity of the critical Coulomb energies 
to the cluster size used in the ED/DMFT, we show in Fig.~\ref{ZU} 
the quantities $Z_i(U)=1/[1- {\rm Im}\,\Sigma_i(i\omega_0)/\omega_0]$
for $n_s=6$ and $n_s=8$ at $T=10$~meV, assuming $J'=J=U/4$.
Here $\Sigma_i(i\omega_0)$ are the subband self-energies at the 
first Matsubara frequency. Thus, for isotropic Hund's 
exchange and $n_s=8$, $U_{c1}\approx2.1$~eV and $U_{c2}\approx3.1$~eV.
(In our notation the $U_{ci=1,2}$ refer here to the subband critical
Coulomb energies for increasing $U$ and should not be confused with 
the stability boundaries for increasing vs decreasing $U$ at the 
individual Mott transitions.) Near both critical Coulomb
energies $Z_i(U)$ exhibit hysteresis behavior characteristic of 
first-order transitions. For $n_s=6$, the common metallic phase 
is stable only up to about $U_{c1}\approx2$~eV.
Thus, inclusion of zero-energy bath levels supports the metallic 
character of the DMFT solution. A shift of about 0.2~eV is found 
for $U_{c2}$ where the wide band becomes insulating. These shifts 
are consistent with single-band ED/DMFT results\cite{prl05} upon 
increasing $n_s$ from 3 to 4; only minor additional shifts occur 
in this case between $n_s=4$ and the fully converged results for 
$n_s=6$.

According to the QMC/DMFT results discussed in Ref.~\cite{prb70}, 
the Mott transitions for $J=U/4$ and $J'=0$, i.e., for Ising-like 
exchange, are located at $U_{c1}\approx2.1$~eV and 
$U_{c2}\approx2.7$~eV. Moreover, apart from the shift of $U_{c2}$, 
this transition is no longer first-order. The two-band ED/DMFT 
results in Ref.~\cite{prl05} for $n_s=6$ confirmed this fundamental 
difference between the $J'=J$ and $J'=0$ treatments.  

As will be discussed in detail below, the wide band in the intermediate
phase, i.e., for $U_{c1}<U<U_{c2}$, does not satisfy Fermi-liquid 
criteria. Thus, the fact that $Z_2(U)$ as defined above is finite 
in this region does not imply existence of ordinary quasiparticles.
In the case of full Hund's coupling, the imaginary part of the 
self-energy of the wide band vanishes in the low-frequency limit, 
but does not increase linearly at small $i\omega_n$, implying 
non-quadratic variation at real frequencies. For Ising-like exchange,
Im\,$\Sigma_2(i\omega_n)$ remains finite for $\omega_n\rightarrow0$,
implying finite lifetime even for states at $E_F$. Plots like those 
in Fig.~1 are nevertheless useful since they permit a convenient
identification of phase transitions. Thus, $Z_i(U)\rightarrow0$ 
at $U_{ci}$ indicates that the narrow or wide subbands become 
insulating for $U>U_{ci}$, respectively. (Of course, due to the 
discrete representation along the imaginary frequency axis at finite 
$T$, $Z_i(U)$ does not fully vanish.) 

\begin{figure}[t!]
  \begin{center}
  \includegraphics[width=5.0cm,height=8cm,angle=-90]{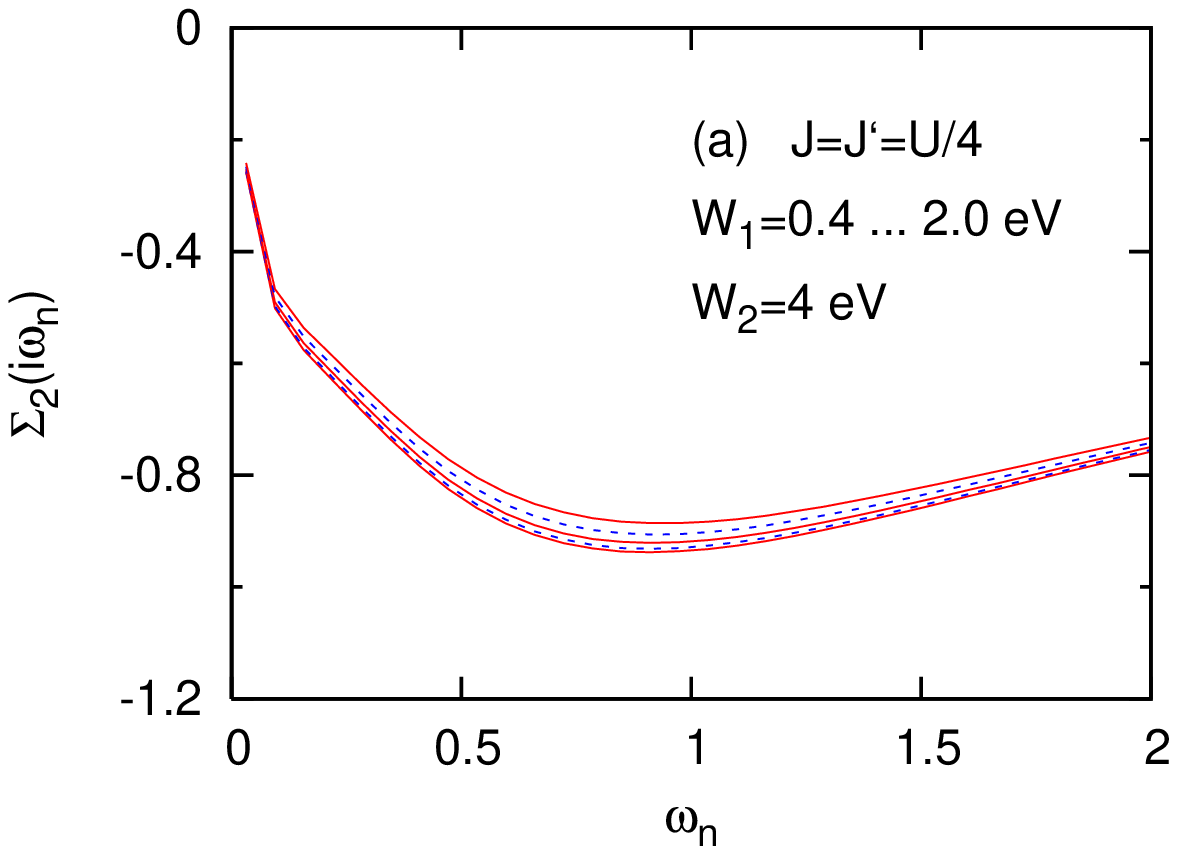}
  \includegraphics[width=5.0cm,height=8cm,angle=-90]{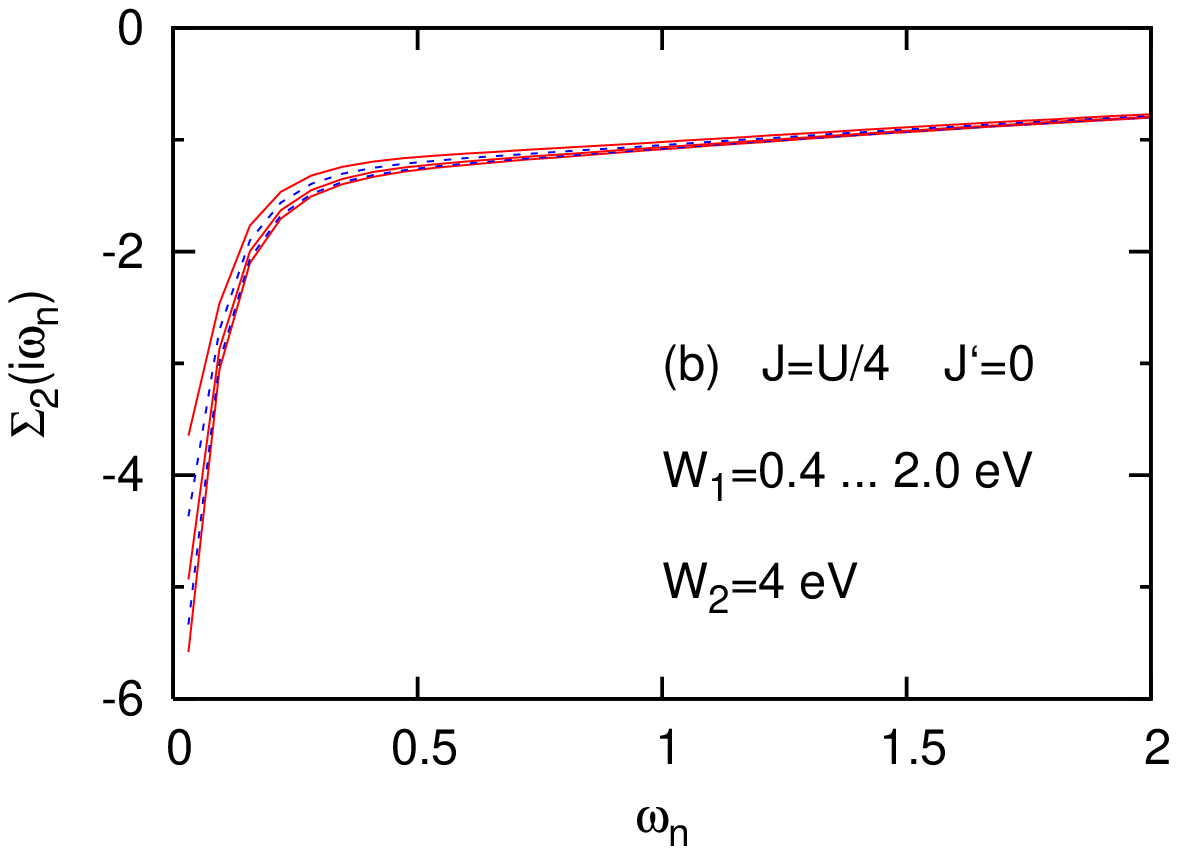}
  \end{center}
  \vskip-2mm
\caption{\label{W1W2}
Imaginary part of self-energy of wide band in intermediate phase
for decreasing widths $W_1$ of narrow band, with fixed $W_2=4$~eV, 
$U=2.4$~eV, $n_s=8$, $T=10$~meV. Alternating solid (red) and dashed
(blue) curves: $W_1=0.4,\,0.8,\,\,1.2,\,1.6,\,2.0$~eV (from bottom).
(a) $J=J'=U/4$; (b)  $J=U/4$, $J'=0$.
}\end{figure}

To check the convergence of the ED results with cluster size $n_s$
we compare them with NRG calculations for an effective one-band model
which is most suitable for the intermediate phase and which is
specified in the following section. Since the narrow band is insulating 
in this phase one of the key features of the effective model is the 
omission of one-electron hopping in the narrow band. Accordingly, the
upper and lower Hubbard peaks will be centered at about $\pm U/2$, but
the influence of their width is neglected. (In a local description, 
the width of the Hubbard bands is approximately given by $W$ \cite{rmp}.)
In Fig.~\ref{W1W2} we show that this approximation indeed has only a 
minor effect on the electronic properties of the metallic wide band. 
Keeping $W_2=4$~eV fixed and changing $W_1$ between $0.1\,W_2$ and 
$0.5\,W_2$, we notice that, both for isotropic and anisotropic exchange 
coupling, the self-energy of the wide band at $U=0.6\,W_2=2.4$~eV
is almost unaffected by the original single-particle width of the 
insulating narrow band. The detailed electronic properties of these 
phases will be discussed in Sections 5 and 6.    

\section{3 \ Numerical renormalization group approach for 
          effective one-band model}

In the limit of an insulating narrow band interacting with a metallic 
wide band, the two-band model can be simplified to an effective one-band 
model by eliminating the high energy states associated with the upper 
and lower Hubbard bands of the narrow band \cite{biermann05}.
Neglecting the one-electron hopping in the narrow band, i.e., assuming 
$W_1=0$, and fixing the occupations of these orbitals, the terms in 
Eq.~(1) involving interorbital Coulomb and pair-exchange interactions
disappear. The effective Hamiltonian then reduces to 
\begin{eqnarray}
  H  & = & \sum_{lm\sigma}t_{lm}c^{+}_{l\sigma}c_{m\sigma}
     + U\sum_l n_{l\uparrow} n_{l\downarrow}\nonumber\\ 
     &&\  - \sum_{l}[2 J S^{z}_{l}s^{z}_{l} +
      2J' (S^{+}_{l}s^{-}_{l}+S^{-}_{l}s^{+}_{l})]\, . \label{eq:fkl}
\end{eqnarray}
In this model, the low energy degrees of freedom of the narrow band 
are represented by local moments $\vec{S}_{l}$. These interact 
ferromagnetically with the local spin density $\vec{s}_{l}$ of the 
wide-band conduction electrons which are also subject to a local 
Coulomb repulsion $U$. The model corresponds to a double-exchange 
model with Coulomb interactions amongst the itinerant electrons or, 
equivalently, to the ferromagnetic Kondo lattice model with interactions 
in the band. The antiferromagnetic case, $J<0$, for $U=0$, has already 
been investigated in the context of heavy fermions  \cite{costi.02}.
We adapt these calculations to the case of interest here, namely,
ferromagnetic exchange, $J>0$, and $U>0$. The equivalence between the 
full two-band model and the effective model holds provided $U$ is large
enough so that a description of the low energy degrees of freedom of 
the narrow band in terms of local moments is possible and provided
that $J' \ll U/2$. The latter condition corresponds to eliminated
excited states of the full model being far from the first excited 
state of the effective model. Unless $J$ is chosen to be very small, 
we see that for realistic values of $J\approx U/4$,
the equivalence will hold well for Ising exchange coupling ($J'=0$) 
but less well for isotropic Hund's coupling ($J'=J$). In either case,
the effective model should be increasingly accurate in the limit 
$\omega, T \rightarrow 0$.


Within DMFT, we need to solve an effective quantum impurity model 
corresponding to a $S=1/2$ Kondo impurity coupled ferromagnetically 
with conduction electrons subject to a local Coulomb repulsion:
\begin{eqnarray}
  H  & = & \sum_{k,\sigma}\varepsilon_{k}c^+_{k\sigma}c_{k\sigma}
  + U n_{0\uparrow}n_{0\downarrow}\nonumber\\ 
  &&\   - 2J S_0^{z}s^{z}_{0} -
      2J' (S_0^{+}s^{-}_{0}+S_0^{-}s^{+}_{0})\, . \label{eq:quantum-imp}
\end{eqnarray}
We solve this model using the numerical renormalization 
group method \cite{wilson.74}, which allows
calculation of dynamical quantities on the real energy axis at both zero
and finite temperature \cite{costi.94,hofstetter.00}.
The impurity self-energy
$\Sigma(\omega)$ is calculated using the method
described in Ref.~\cite{bulla.98}. For comparison with ED, it is then 
evaluated on the imaginary axis $z=i\omega$ by analytic continuation
\begin{eqnarray}
 \Sigma(i\omega) = -\frac{1}{\pi}\int_{-\infty}^{+\infty} d\omega'\ 
    \frac{{\rm Im}\,\Sigma(\omega')}{i\omega-\omega'}\, .
\label{eq:matsubara-se}
\end{eqnarray}
The calculations use a logarithmic discretization of the conduction band 
$\varepsilon_{k_n}\rightarrow \pm D_2 \Lambda^{-\frac{n-1}{2}}$ with
$\Lambda=1.5$ and we retain of order 600 states per NRG iteration. 
Details of the calculation of spectra and other dynamical quantities 
can be found in Refs.~\cite{costi.94,bulla.01}.

The effective model specified in Eq.~(\ref{eq:fkl}) 
allows the non-Fermi-liquid physics of the intermediate phase of the 
two-band model to be understood qualitatively \cite{biermann05}. 
The case of Ising exchange ($J'=0$) describes interacting 
conduction electrons scattering from a disordered static configuration 
of local spins with $S_{l}^{z}=\pm 1/2$. The disorder potential
is proportional to the Ising exchange $J$. The conduction electrons
therefore have a finite lifetime at the Fermi level even at $T=0$, 
as is characteristic of a disordered metal. Switching on the
spin-flip part of the Hund's exchange gives the disordered spin 
configuration some dynamics at finite temperature, but as long as 
$J' < J$, we expect from the ferromagnetic Kondo model that the 
spin-flip part of the Hund's exchange will renormalize to zero 
at low temperature with the Ising part of the Hund's exchange remaining
finite. At $T=0$, the system will again be disordered with a disorder 
potential given by the finite Ising part of the Hund's exchange. 
This results in a finite lifetime for the conduction electrons at $T=0$. 

The situation changes for isotropic Hund's exchange ($J'=J$). 
In this case, the ferromagnetic exchange in the effective impurity 
model (\ref{eq:quantum-imp}) is known to be marginally irrelevant 
\cite{cragg.79,koller.05}. 
Thus, both Ising and spin-flip parts of the Hund's exchange renormalize
to zero and there will be no disorder scattering at $T=0$, giving rise
to an infinite lifetime for the conduction electrons at $T=0$. The vanishing
of the self-energy at $\omega=T=0$ also implies that the narrow-band
spectral function satisfies a pinning condition at $\omega = T =0$.
In Sections 5 and 6, these qualitative considerations will be seen 
to be in accord with the numerical results from ED for the full 
two-band model and NRG results for the effective one-band model.

\section{4 \ Metallic phase}

We begin our discussion of the multi-band ED results with the region 
in which both subbands are metallic, just below the Mott transition 
associated with the narrow band. To illustrate the applicability of 
the ED approach this region is very useful since it involves 
important redistribution of spectral weight between low and high 
frequencies that must be captured properly in order to describe
the Mott transition.

\begin{figure}[t!]
  \begin{center}
  \includegraphics[width=5.0cm,height=8cm,angle=-90]{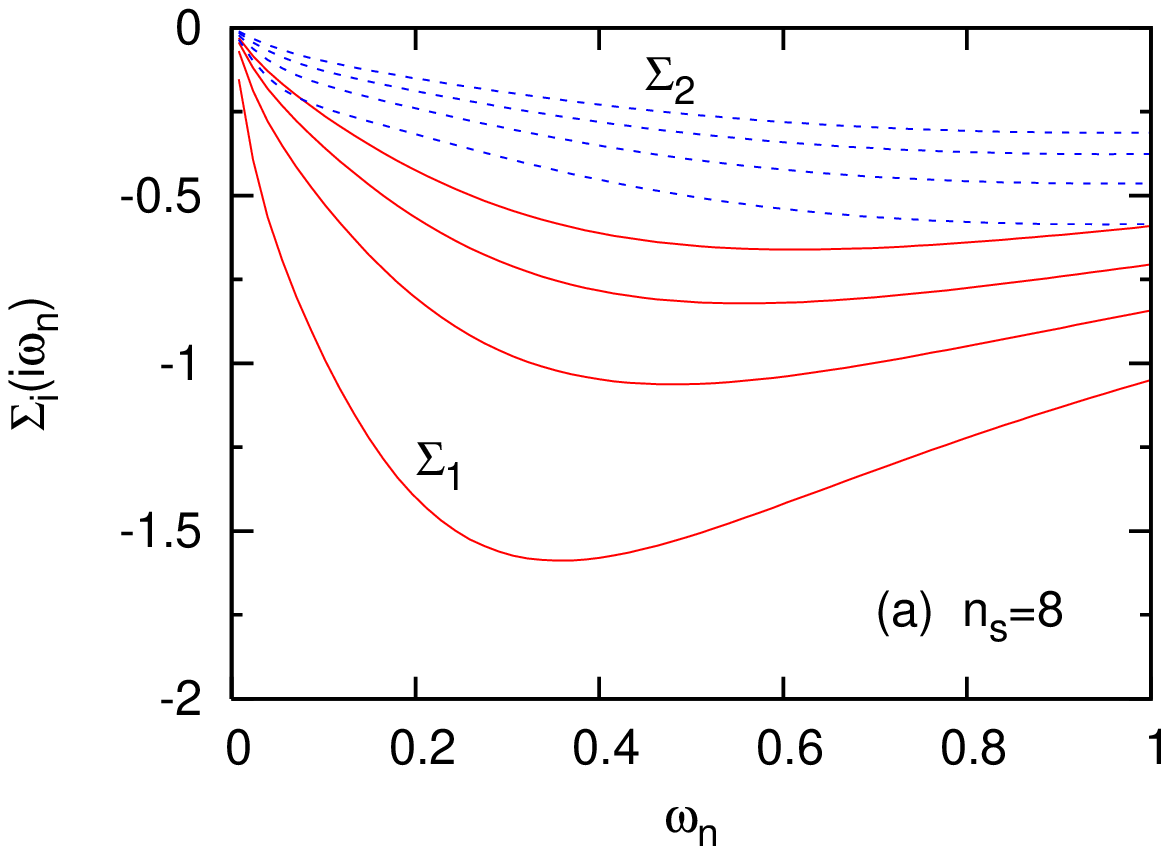}
  \includegraphics[width=5.0cm,height=8cm,angle=-90]{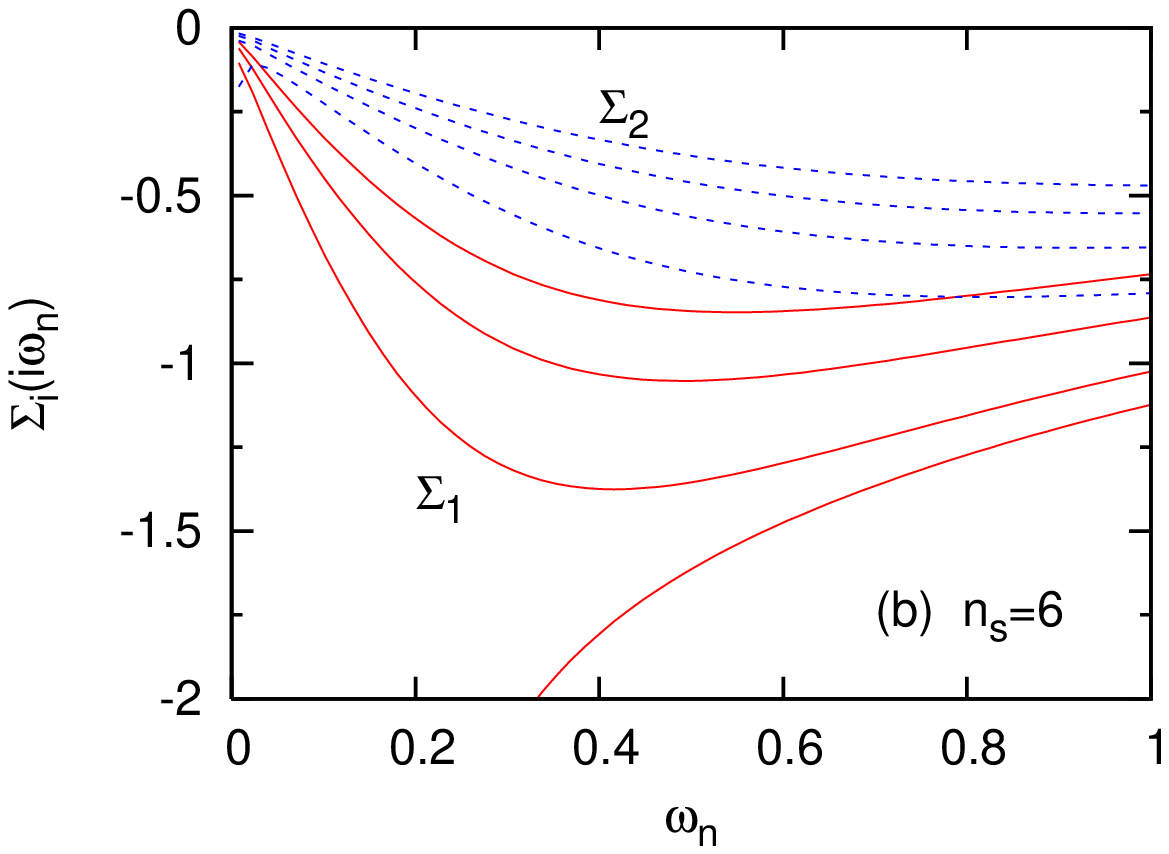}
  \end{center}
  \vskip-2mm
\caption{\label{M8M6}
Imaginary part of subband self-energies in metallic region 
for $J=J'=U/4$ at $T=2.5$~meV. 
Solid (red) curves: narrow band, dashed (blue) curves: wide band. 
From top: $U=1.8,\ 1.9,\,2.0,\,2.1$~eV.   (a) $n_s=8$; (b) $n_s=6$.
}\end{figure}

Fig.~\ref{M8M6} shows a comparison of the subband self-energies for 
$n_s=8$ and $n_s=6$ in the region close to the lower critical Coulomb 
energy for $T=2.5$~meV. For $n_s=8$, the $\Sigma_i(i\omega_n)$ vary 
roughly linearly at low frequencies, indicating that both bands are 
metallic. Of course, the narrow band is more strongly correlated than 
the wide band. The spacing between the lowest excited state and the 
ground state energy in these calculations is typically $1-2$~meV, 
i.e., evaluation of the self-energies in this low temperature range 
is indeed meaningful.

The results for $n_s=6$ are similar, except that the self-energy of the 
narrow band at $U=2.1$~eV is inversely proportional to $i\omega_n$,
indicating that the Mott transition in this case is located between
$U=2.0$~eV and $U=2.1$~eV. In addition, at the lowest Matsubara 
frequencies the self-energy of the wide band at 2.1~eV exhibits  
deviations from linear $i\omega_n$ variation. These deviations 
are related to finite-size effects and will be analyzed in more 
detail in the next section when we discuss the intermediate 
orbital-selective phase. Just below the transition for $n_s=6$, at  
$U=2$~eV, very small deviations from approximately linear $i\omega_n$ 
variation can also be seen in both $\Sigma_i(i\omega_n)$. 

It is remarkable that the ED results for $n_s=8$ capture the correlation 
effects in the common metallic phase of both subbands very well until 
close to $U_{c1}$ and down to rather low temperatures. This result 
is not at all obvious since so close to the Mott transition a large 
fraction of the spectral weight of the narrow band is transfered from 
the Fermi level to the Hubbard bands. Even the results for $n_s=6$ are  
qualitatively correct. The main effect of the cluster size $n_s=6$ is 
the underestimate of $U_{c1}$ by about $0.1$~eV. According to the 
single-band results \cite{prl05} inclusion of the zero energy bath levels 
in the $n_s=8$ cluster should provide the most important part of the 
shift towards the correct $U_{c1}$. In addition these extra bath levels 
yield an excellent representation of the metallic properties of both 
subbands right up to the critical Coulomb energy. 

Of course, slight finite-size effects should be manifest also in the 
low-temperature ED results for $n_s=8$. The precise form of  
$\Sigma_i(i\omega_n)$ at very low frequencies and temperatures,
in particular, the range of true Fermi-liquid behavior in the immediate 
vicinity of $U_{c1}$, can only be investigated by more accurate methods, 
such as a full two-band extension of the NRG approach which is applicable 
at zero and finite $T$ \cite{costi}. 

We emphasize that this region of the $T/U$ phase diagram for isotropic 
Hund's exchange is not yet accessible using QMC calculations. 
It would therefore be of great interest to extend the present results 
to $n_s=9$ and $n_s=12$ in order to explore the strongly correlated 
metallic phase of realistic three-band materials which have so far been 
investigated only for Ising-like exchange interactions. A more detailed 
comparison of the role of Hund vs. Ising exchange treatments in the 
metallic phase of the present two-band model will be given elsewhere
\cite{metallic}.     

\section{5 \ Intermediate phase: isotropic Hund's exchange}

As shown first by Koga {\it et al.}~\cite{koga1} within ED/DMFT 
calculations at $T=0$, the two-band Hubbard model with full Hund's exchange 
interaction exhibits successive, orbital-selective Mott transitions. 
In Ref.~\cite{prl05} we proved that analogous ED calculations at 
finite temperature are consistent with these findings, revealing 
sequential first-order transitions associated with the two subbands. 
In this section we discuss in more detail the low temperature properties 
of the intermediate phase between the fully metallic and insulating phases.

\begin{figure}[t!]
  \begin{center}
  \includegraphics[width=5.0cm,height=8cm,angle=-90]{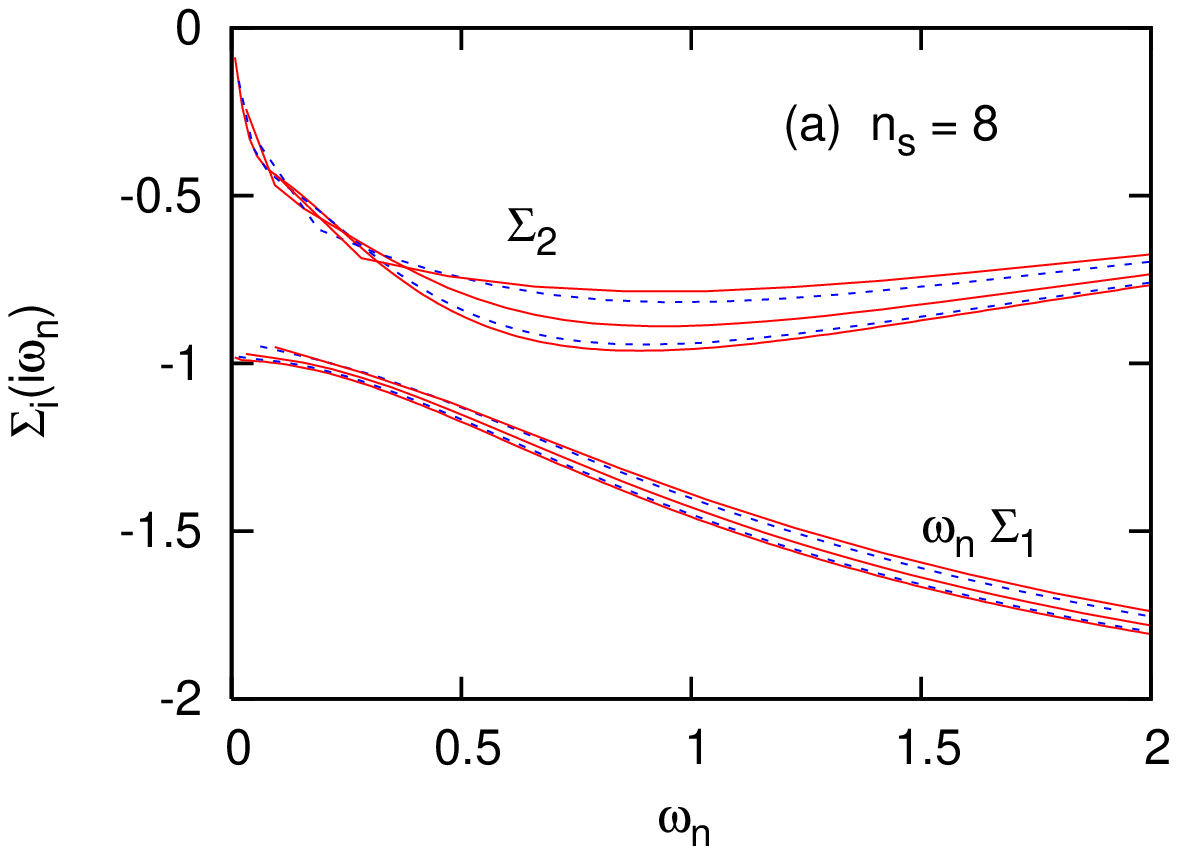}
  \includegraphics[width=5.0cm,height=8cm,angle=-90]{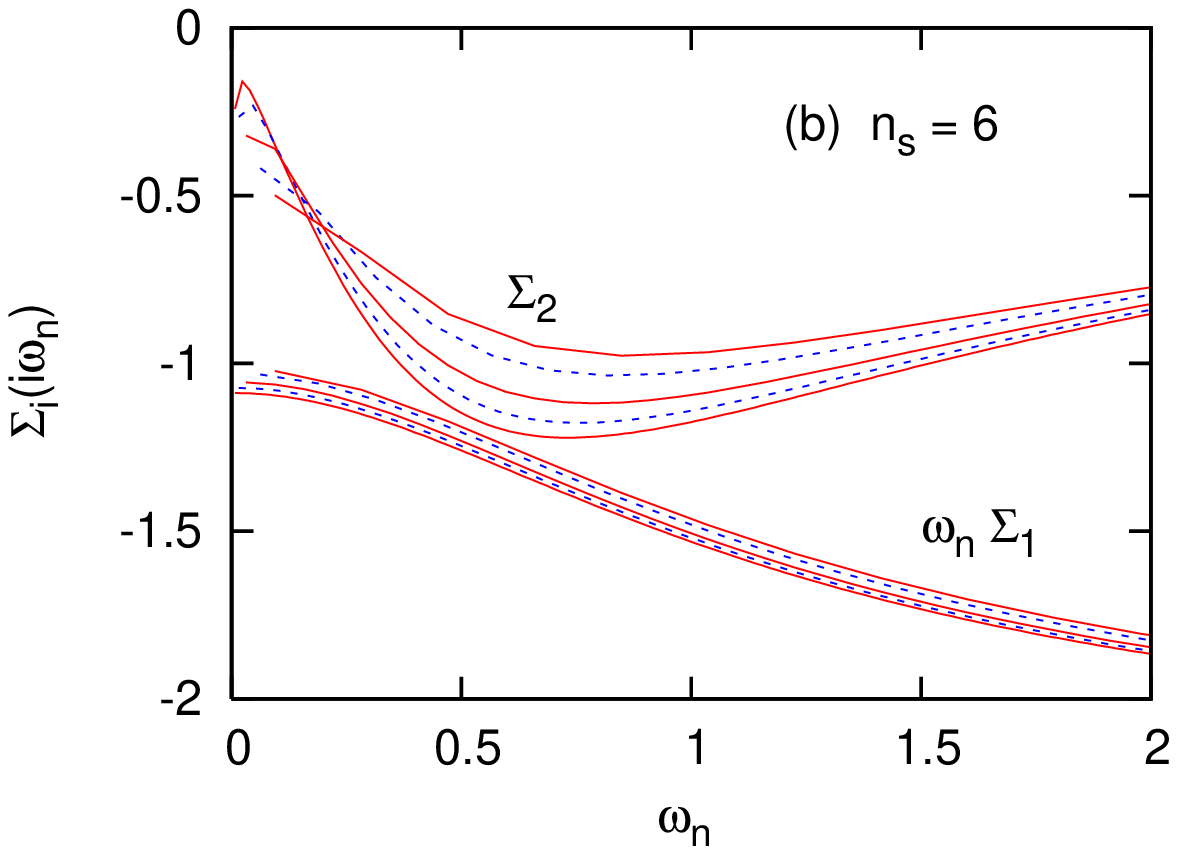}
  \end{center}
  \vskip-2mm
\caption{\label{OSMT}
Imaginary part of subband 
self-energies in intermediate phase, for $U=2.4$~eV and $J=J'=U/4$.  
$T=2.5,\ 5,\ 10,\ 20,\ 30$~meV (from bottom). 
(a) $n_s=8$; (b) $n_s=6$; 
}\end{figure}

Fig.~\ref{OSMT} shows the subband self-energies for a Coulomb energy 
$U=2.4$~eV, i.e., between the lower and upper Mott transitions. 
The results 
for $n_s=8$ demonstrate that the narrow band is insulating, i.e., its
self-energy is inversely proportional to $i\omega_n$ at low frequencies. 
$\Sigma_1(i\omega_n)$ is also nearly independent of temperature in the 
range $T=2.5\ldots 30$~meV, suggesting that the Mott gap is considerably 
larger. The self-energy of the wide band, however, extrapolates to zero 
at low frequencies, indicating metallic behavior. These results 
suggest that the pinning condition $N_2(0)=A_2(0)$ is satisfied, i.e., 
the spectral density at $E_F$ of the interacting system coincides 
with that of the non-interacting system. Because of the rapid increase 
of Im\,$\Sigma_2(i\omega_n)$ at small $\omega_n$, the spectral 
distribution of the wide band in the intermediate phase consists of 
a very narrow peak at $E_F$ and large upper and lower Hubbard bands.
Nevertheless, as will be discussed further below, the comparison 
with the NRG results indicates that this band does not satisfy 
true Fermi-liquid behavior.

The results for $n_s=6$ shown in Fig.~\ref{OSMT}(b) agree qualitatively 
with those for $n_s=8$. The self-energies of the narrow band are nearly 
identical, with only a slightly larger dependence on temperature in 
the case $n_s=6$. The main difference is that $\Sigma_2(i\omega_n)$ 
at low $T$ and small $i\omega_n$ exhibits larger deviations as a result 
of finite-size effects. This is plausible since, as discussed above, 
the wide band is metallic, but highly correlated. The pronounced 
three-peak structure of its spectral function is therefore not 
represented accurately by only two non-zero bath levels per 
impurity orbital in the ED calculation for $n_s=6$. The extra bath
level with zero energy included for $n_s=8$ therefore plays a crucial 
role for a proper description of the metallic behavior of the wide
band for full Hund's coupling.   

\begin{figure}[t!]
  \begin{center}
  \includegraphics[width=5.0cm,height=8cm,angle=-90]{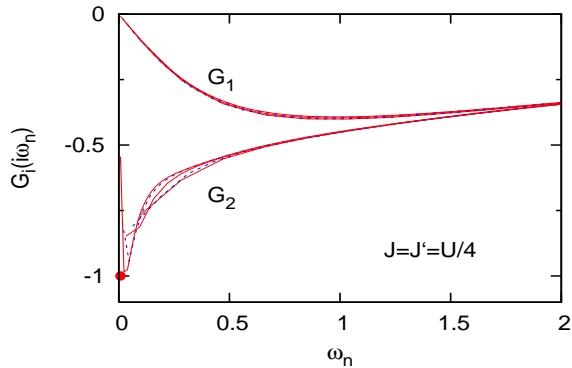}
  \end{center}
  \vskip-2mm
\caption{\label{GreenJ}
Imaginary part of subband 
Green's functions in intermediate phase, for $U=2.4$~eV, $J=J'=U/4$,
$n_s=8$. $T=2.5,\ 5,\ 10,\ 20,\ 30$~meV. 
The dot indicates the pinning condition for the wide band.
}\end{figure}

Fig.~\ref{GreenJ} shows the subband Greens functions $G_i(i\omega_n)$ 
in the intermediate phase at various temperatures. Im\,$G_1$ is linear in 
$\omega_n$ at low frequencies, as expected for the insulating subband. 
Although the wide band exhibits clear signs of finite-size effects at 
small $\omega_n$ it is nevertheless seen to approximately satisfy the 
pinning condition Im\,$G_2(i\omega_n)\rightarrow -\pi N_2(0) = -1$ in 
the limit $\omega_n\rightarrow0$.   

For computational reasons the role of finite-size effects in the 
present finite-temperature ED approach based on full matrix 
diagonalization can at present not be checked by going beyond the 
cluster size $n_s=8$. To analyze the low-frequency properties of 
the wide band in the intermediate phase more closely, we have 
carried out NRG calculations for the effective one-band model  
discussed in Section 3. Although this model is more appropriate
for the anisotropic exchange treatment discussed in the next section,
we use it here (with some caution) since it is so far the only scheme 
capable of providing a guideline for the two-band Hubbard model 
at low finite temperatures.

\begin{figure}[t!]
  \begin{center}
  \includegraphics[width=5.0cm,height=8cm,angle=-90]{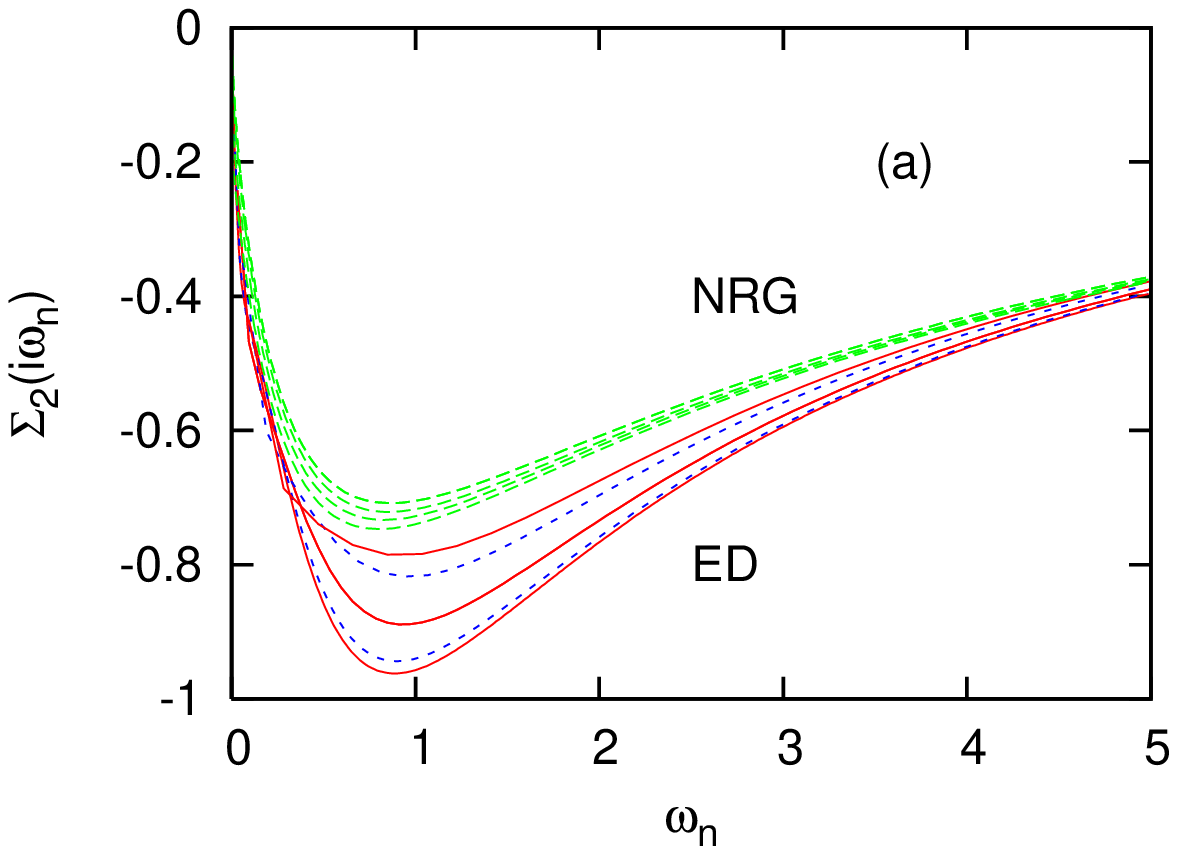}
  \includegraphics[width=5.0cm,height=8cm,angle=-90]{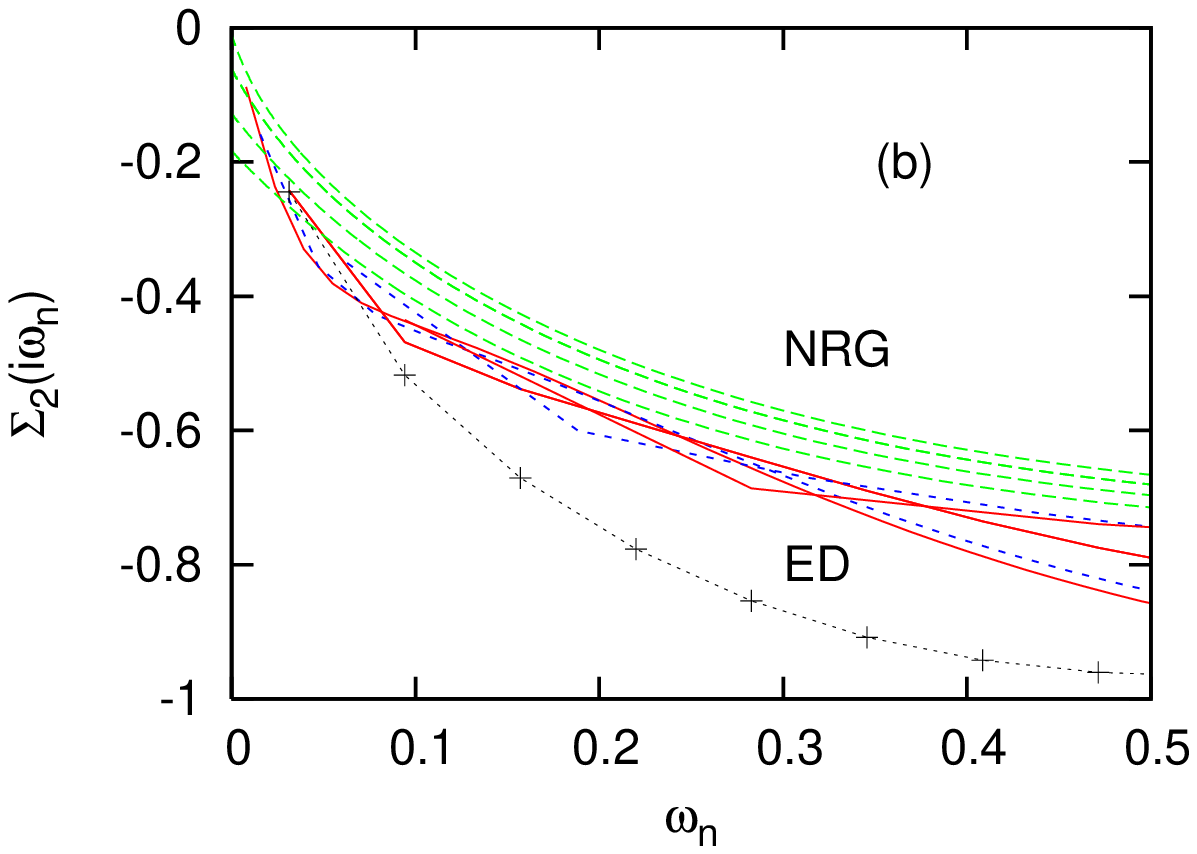}
  \end{center}
  \vskip-2mm
\caption{\label{NRG34}
Imaginary part of self-energy $\Sigma_2$ of wide band for $U=2.4$~eV 
and $J=J'=U/4$. (a) Solid (red) and dashed (blue) curves: ED results 
for $n_s=8$ at  $T=2.5,\,5,\,10,\,20,\,30$~meV (from bottom); 
green curves: NRG results for $T=3,\ 10,\ 15,\ 34$~meV (from top). 
(b) Same quantities on expanded low-frequency scale. Black curve (+):
self-energy $\Sigma_1$  of narrow band at $2.0$~eV. 
}\end{figure}

In Fig.~\ref{NRG34}(a) the ED results for Im\,$\Sigma_2(i\omega_n)$ are 
compared with the corresponding self-energy derived within the NRG. 
The overall frequency variation of $\Sigma_2$ is seen to be remarkably 
similar for both methods, implying similar spectral distributions.
The most noticeable difference is the larger dependence on temperature 
in the case of the ED results.
According to Fig.~\ref{OSMT}, the increase in cluster size from $n_s=6$ 
to $n_s=8$ diminishes the variation of $\Sigma_2$ with temperature and 
makes the minimum near $\omega_n\approx0.9$ less deep. This trend 
indicates that larger cluster sizes would bring the ED results
into better agreement with the NRG data. We point out, however, 
that part of the difference should be caused by the approximate 
nature of the NRG model in the case $J'=J$. 
As shown in the following section, for $J'=0$ the ED and NRG results 
for $\Sigma_2$ at $T=20\ldots30$~meV are in perfect agreement.  

The low-frequency region of $\Sigma_2(i\omega_n)$ is plotted in 
greater detail in Fig.~\ref{NRG34}(b). In the ED calculation, the 
spacing between the lowest excited level and the ground state energy 
is typically $10$~meV. Thus, temperatures below this range will give
rise to increasing finite-size effects, precluding any meaningful 
analysis at low frequencies. Despite this limitation, the ED data 
follow the frequency dependence of the NRG results rather well. 

In order to provide some insight into the variation of $\Sigma_2$ 
at finite $\omega_n$, we compare it in Fig.~\ref{NRG34}(b) also 
with the self-energy of the narrow band  $\Sigma_1$
at $U=2.0$~eV, i.e., 
just below the Mott transition at $U_{c1}\approx2.1$~eV. 
According to Fig.~1(a), we find: 
$Z_1(U=2.0~{\rm eV})\approx Z_2(U=2.4~{\rm eV})\approx 0.11$. 
Nevertheless, whereas $\Sigma_1$ at $U=2.0$~eV exhibits the typical 
behavior expected for a strongly correlated Fermi-liquid phase, 
$\Sigma_2$ evidently has much larger nonlinear corrections. A similar 
systematic difference is found between $\Sigma_2$ in the 
intermediate phase and the behavior the self-energy in a 
single-band model for $U$ just below the Mott transition.

The NRG results shown in Fig.~\ref{NRG34} indicate that   
$\Sigma_2(i\omega_n)\rightarrow0$ in the limit $\omega_n\rightarrow0$ 
and for low $T$. Thus, particles at $E_F$ have infinite lifetime. 
In addition, analogous NRG calculations at $T=0$ yield \cite{costi}  
non-Fermi-liquid behavior, in agreement with Ref.~\cite{biermann05}.
Thus, at small finite $i\omega_n$ Im\,$\Sigma_2$ does not vary 
linearly. The qualitative agreement between the finite-$T$ NRG and 
ED results seen in Fig.~\ref{NRG34} shows that the 
ED data are consistent with non-Fermi-liquid behavior.  

\begin{figure}[t!]
  \begin{center}
  \includegraphics[width=5.0cm,height=8cm,angle=-90]{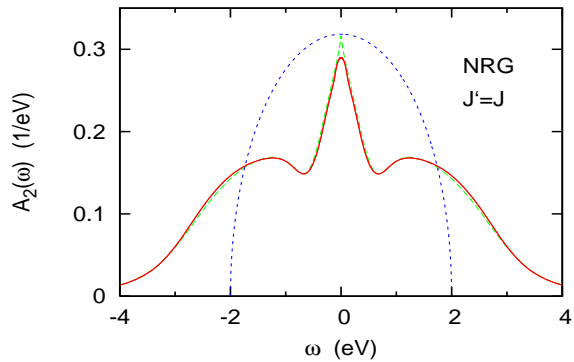}
  \end{center}
  \vskip-2mm
\caption{\label{NRG12}
Spectral distribution of wide band in intermediate phase,
calculated within NRG approach for $U=2.4$~eV, $J=J'=U/4$. 
Green curve: $T=3$~meV; red curve: $T=30$~meV;  
dashed blue curve: bare density of states.
}\end{figure}
    
Fig.~\ref{NRG12} shows the spectral function of the wide band in the
intermediate phase, calculated within the NRG. Spectra of this kind 
were also obtained by Arita and Held \cite{arita} for $U$ near $U_{c1}$ 
within QMC/DMFT calculations for the same two-band model for $T=0$ and
$J'=J=U/4$.
As mentioned above, the distribution exhibits a very narrow peak which 
satisfies the pinning condition at $E_F$ for temperatures up to about 
10~meV. 
Because of the sharp central peak of $A_2(\omega)$, the real and 
imaginary parts of the self-energy of the wide band $\Sigma_2(\omega)$ 
exhibit structure on a similar scale. Nevertheless, the variation 
of Im\,$\Sigma_2(i\omega_n)$ obtained in the NRG is very smooth. 
Thus, the weak shoulder seen in the ED results for $\Sigma_2$ near 
$\omega_n\approx0.06$ in Figs.~\ref{W1W2}(a), \ref{OSMT}(a) and 
\ref{NRG34}  must be attributed to finite-size effects.   

The ED and NRG results discussed above demonstrate that, in the 
presence of full Hund's exchange, the wide band above the main 
Mott transition at $U_{c1}$ remains metallic, without satisfying
Fermi-liquid criteria. In the following section it will 
be shown that the absence of spin-flip and pair-exchange terms
enhances this trend towards non-Fermi-liquid behavior, so that 
even particles at $E_F$ acquire a finite lifetime. Thus, in both
cases, the intermediate phase is bad-metallic.  

We point out that the differences between the self-energies for 
$n_s=8$ and $n_s=6$ appear mainly at rather low temperatures, below
about $T=20$~meV. Since QMC calculations also become difficult to 
converge in this range, three-band ED calculations with 
two bath levels per impurity orbital, i.e., cluster size $n_s=9$, 
should be competitive with analogous QMC calculations -- with the 
important additional benefit, that the ED approach can handle 
full Hund's coupling.          

\section{6 \ Intermediate phase: Ising exchange}

As shown in Ref.~\cite{prb70}, in the absence of spin-flip and 
pair-exchange contributions, the first-order Mott transition at 
$U_{c1}\approx 2.1$~eV affects the different subbands of the 
Hubbard model in fundamentally different ways: The narrow band 
undergoes a complete metal insulator transition, but the wide
band changes from a normal metal to a bad metal in the sense
that its self-energy exhibits progressive deviations from 
Fermi-liquid linear $\omega_n$ variation at low frequencies.
Thus, Im\,$\Sigma_2(i\omega_n)\rightarrow c(U)$ for     
$\omega_n\rightarrow 0$, where  $c(U)\approx 0\rightarrow-\infty$ 
for $U = U_{c1}\rightarrow U_{c2}$. Spectral functions obtained via 
the maximum entropy method showed that this breakdown of Fermi-liquid
behavior in the wide band in the intermediate phase leads to a narrow 
pseudogap near $E_F$ and to a violation of the pinning condition, 
i.e., $A_2(0)<N_2(0)$. The spectral function of the wide band then 
acquires a characteristic four-peak structure, with two maxima flanking 
the pseudogap, in addition to the Hubbard bands at higher energies.  
With increasing $U$, the pseudogap becomes deeper and wider, until 
it turns into a true insulating gap at $U_{c2}\approx 2.7$~eV.
As also shown in Ref.~\cite{prb70}, the upper bad-metal to insulator
transition is not a first-order transition, in contrast to the main
transition at $U_{c1}$.

In this section we discuss the temperature dependence of the subband
self-energies in the intermediate phase and the finite-size effects 
associated with the limited number of bath levels included in our 
ED/DMFT approach.

\begin{figure}[t!]
  \begin{center}
  \includegraphics[width=5.0cm,height=8cm,angle=-90]{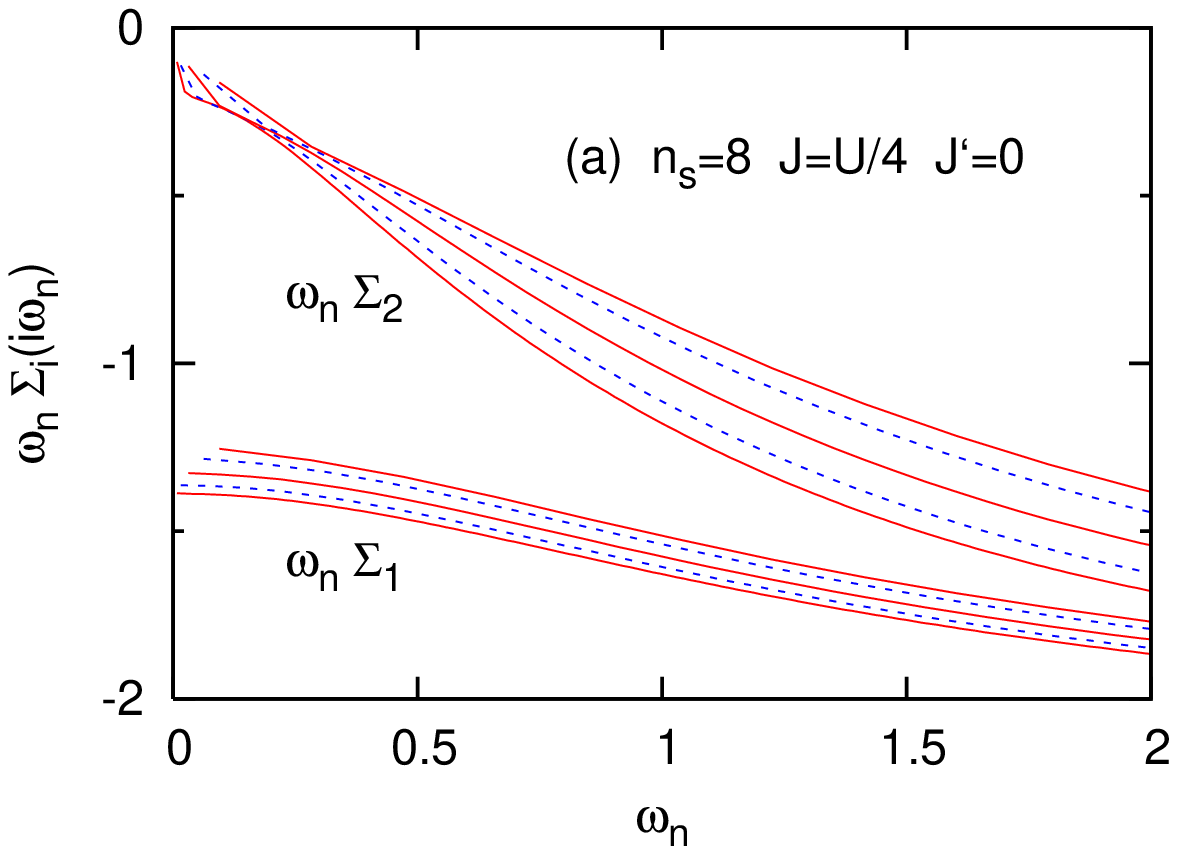}
  \includegraphics[width=5.0cm,height=8cm,angle=-90]{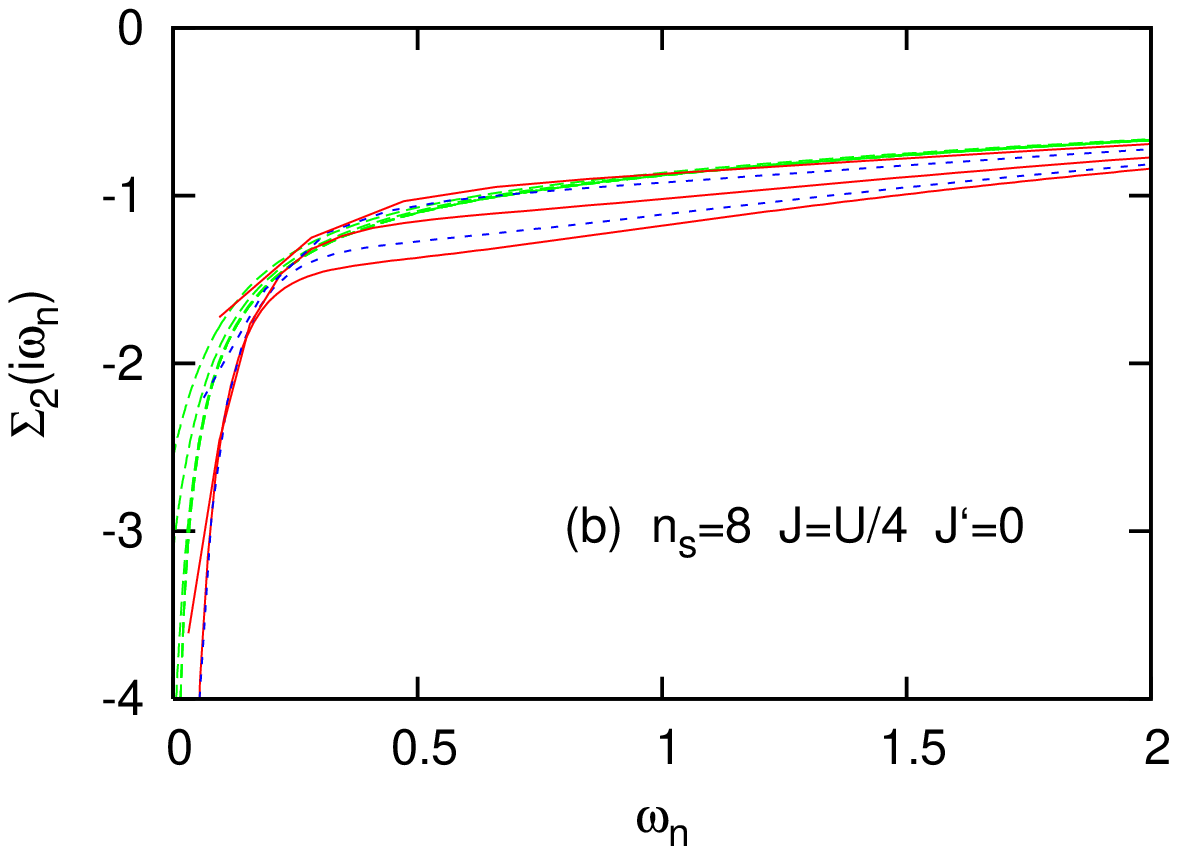}
  \includegraphics[width=5.0cm,height=8cm,angle=-90]{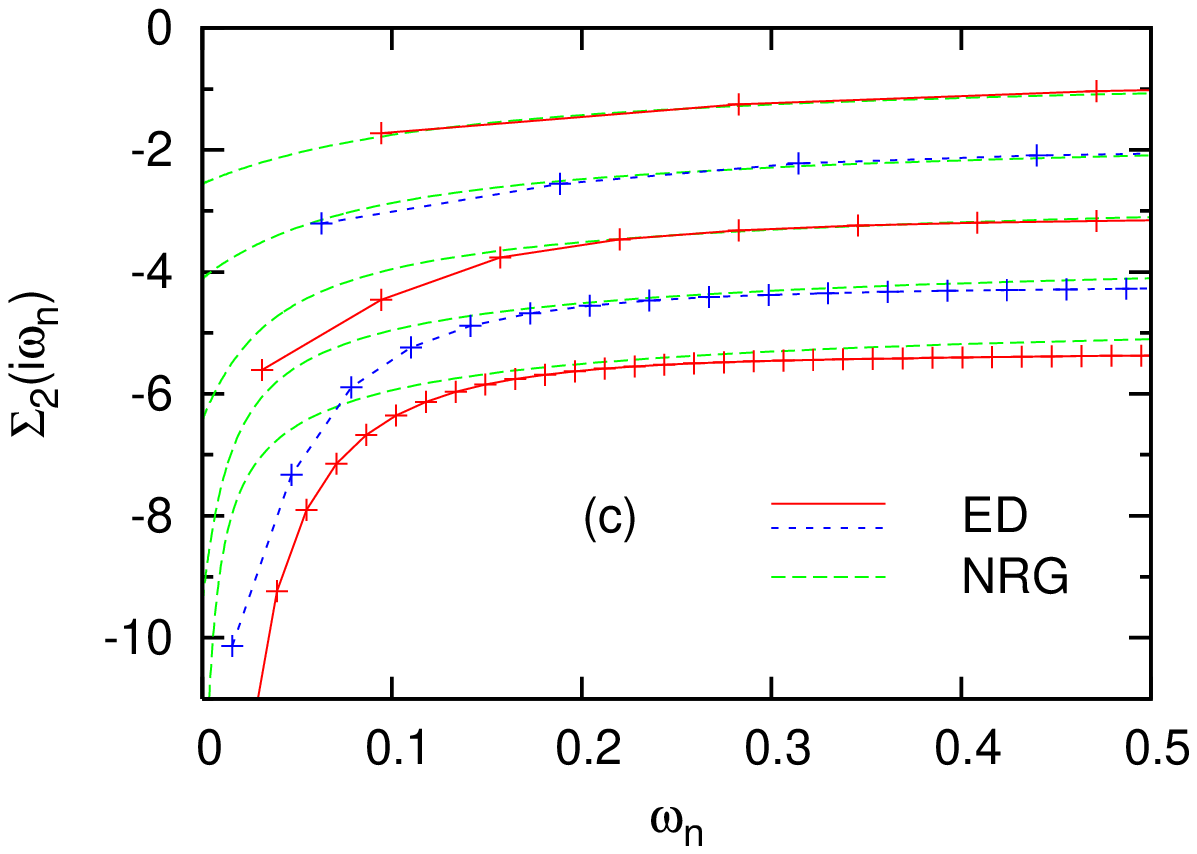}
  \end{center}
  \vskip-2mm
\caption{\label{NRG345J0}
Imaginary part of subband self-energies $\Sigma_i(i\omega_n)$ 
in intermediate phase for $n_s=8$, $U=2.4$~eV, $J=U/4$, $J'=0$;
Solid (red) and dashed (blue) curves:  ED results for    
$T=2.5,\ 5,\ 10,\ 20,\ 30$~meV (from bottom). 
(a)  $\omega_n\Sigma_i(i\omega_n)$;
(b)  $\Sigma_2(i\omega_n)$; green curves: NRG results as in (c).
(c) Low-frequency behavior of self-energy of wide band: 
red and blue curves: ED results as in (b); green curves: NRG results 
for $T=2,\,4.4,\,10,\,20,\,30$~meV (from bottom). 
For clarity, successive curves in (c) are displaced vertically by 1.
}\end{figure}

Fig.~\ref{NRG345J0}
shows the subband self-energies for $n_s=8$ and $U=2.4$~eV,
assuming again $J=U/4$, but $J'=0$. The narrow band is fully insulating, 
so that $\omega_n\,{\rm Im}\,\Sigma_1(i\omega_n)\rightarrow {\rm const.}$ 
nearly independently of temperature, similarly to the case $J'=J$ plotted 
in Fig.~\ref{OSMT}. The self-energy of the wide band, however, differs 
qualitatively from the behavior found for isotropic Hund's coupling. 
Rather than vanishing in the limit $\omega_n\rightarrow0$, 
Im\,$\Sigma_2(i\omega_n)$ now approaches a sharp minimum, which gets 
more pronounced towards low temperatures. Clearly, since    
$\omega_n\,{\rm Im}\,\Sigma_2(i\omega_n)\rightarrow0$ at small 
$\omega_n$, this band is not yet insulating.  
Thus, instead of satisfying the pinning condition at $E_F$, the 
spectral function of the wide band exhibits a dip or pseudogap at
the Fermi level (see below).

Fig.~\ref{NRG345J0}(b) also shows the NRG results for 
$\Sigma_2(i\omega_n)$. 
They are seen to be in very good agreement with the ED data, 
except at low frequencies for $T\le10$~meV. 
The fact that for $T=20\ldots30$~meV there is now much better
coincidence between the ED and NRG results than in Fig.~\ref{NRG34} 
for $J'=J$ indicates that, as discussed in Section 3, the 
effective one-band model employed 
for the NRG is more appropriate in the case $J'=0$. The low-frequency 
behavior of Im\,$\Sigma_2(i\omega_n)$ is shown in more detail 
in Fig.~\ref{NRG345J0}(c). Both the ED and NRG results reveal that the 
sharp minimum of $\Sigma_2$ at small $\omega_n$ gets progressively 
deeper at low temperatures, suggesting that the pseudogap in the spectral
function becomes accordingly deeper.

\begin{figure}[t!]
  \begin{center}
  \includegraphics[width=5.0cm,height=8cm,angle=-90]{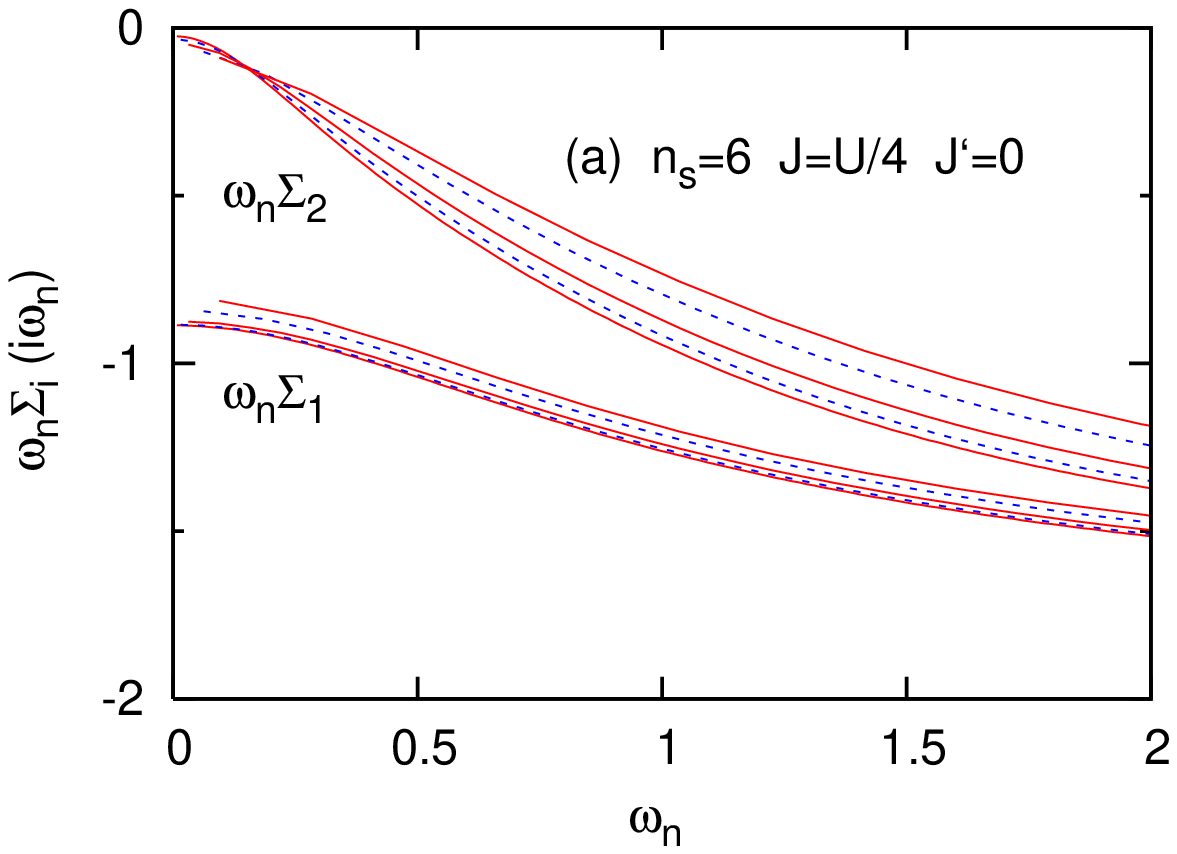}
  \includegraphics[width=5.0cm,height=8cm,angle=-90]{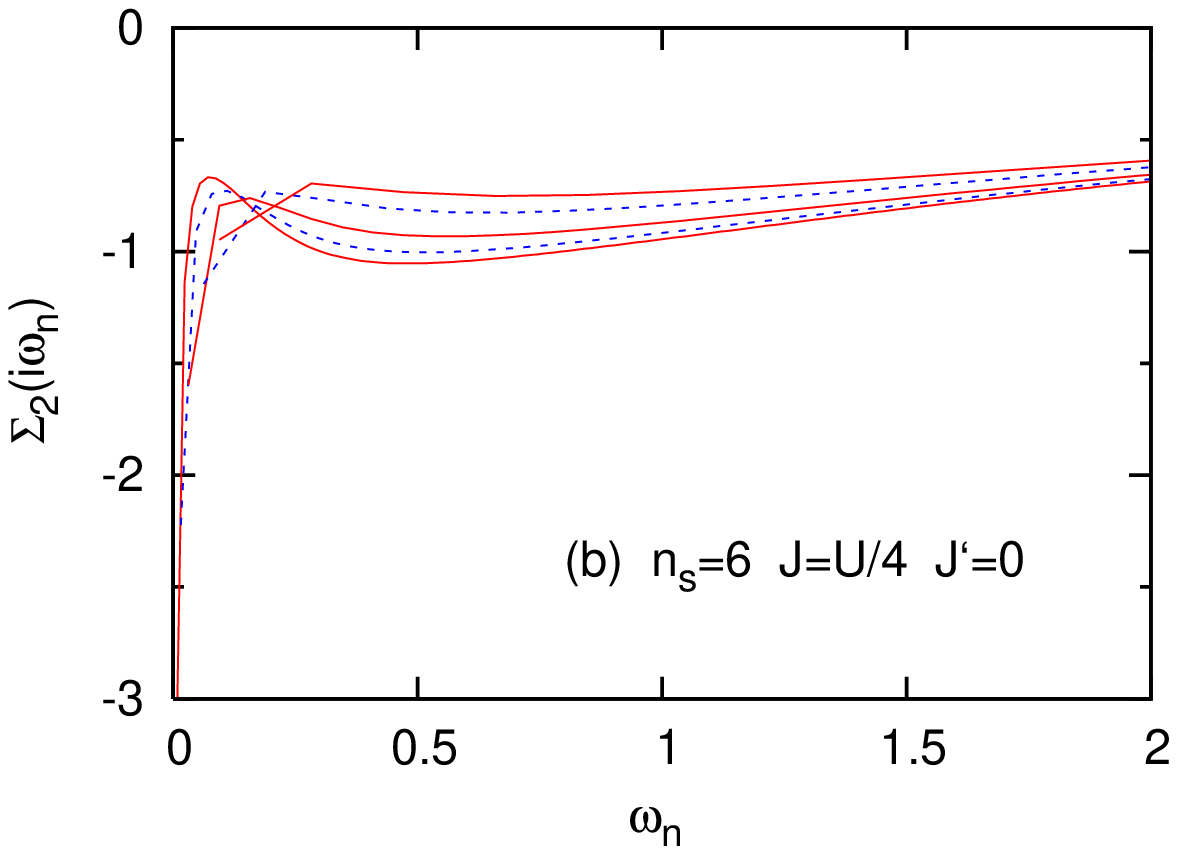}
  \end{center}
  \vskip-2mm
\caption{\label{NFL6}
Imaginary part of subband self-energies $\Sigma_i(i\omega_n)$ 
in intermediate phase for $n_s=6$, $U=2.2$~eV and $J=U/4$, $J'=0$.   
$T=2.5,\ 5,\ 10,\ 20,\ 30$~meV (from bottom).  
}\end{figure}

Analogous results for $n_s=6$ are given in Fig.~\ref{NFL6}. Since 
according to Ref.~\cite{prl05} in this case the wide band becomes 
insulating near $U\approx2.4$~eV, we choose $U=2.2$~eV to illustrate 
the non-Fermi-liquid behavior of this band in the intermediate phase. 
As for $n_s=8$, the narrow band is insulating, i.e., 
$\omega_n\,{\rm Im}\,\Sigma_1(i\omega_n)\rightarrow{\rm const.}$ 
at low frequencies. The frequency variation of $\Sigma_2(i\omega_n)$ 
is seen to be more strongly affected by finite-size effects. 
Nevertheless, as in the case $n_s=8$, a sharp minimum is found 
for $\omega_n\rightarrow0$.

\begin{figure}[t!]
  \begin{center}
  \includegraphics[width=5.0cm,height=8cm,angle=-90]{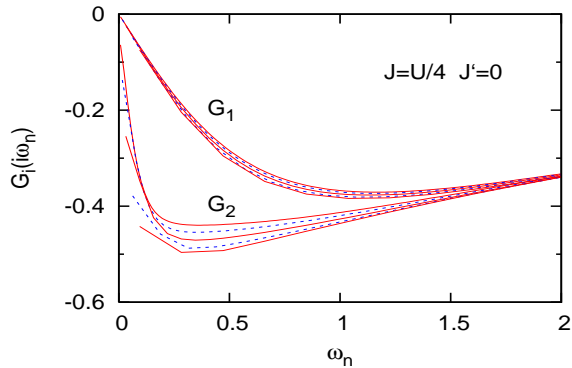}
  \end{center}
  \vskip-2mm
\caption{\label{GreenJ0}
Imaginary part of subband 
Green's functions in intermediate phase, for $U=2.4$~eV, $J=U/4$, 
$J'=0$, $n_s=8$. $T=2.5,\ 5,\ 10,\ 20,\ 30$~meV (from top). 
}\end{figure}

Fig.~\ref{GreenJ0} shows the subband Green's functions $G_i(i\omega_n)$ 
for anisotropic Hund's exchange at various temperatures. 
As for the case $J'=J$ shown in Fig.~\ref{GreenJ}, Im\,$G_1$ is linear 
in $\omega_n$ at low frequencies, since the narrow band is insulating.
On the other hand, because of the more severe breakdown of Fermi-liquid 
behavior in the wide band for $J'=0$, with Im\,$\Sigma_2(i\omega_n)\ne 0$ 
in the low-frequency limit, the Green's function no longer satisfies the 
pinning condition. Thus, Im\,$G_2(i\omega_n)\rightarrow -c(T)$ with 
$c(T)<\pi N_2(0) = 1$, and $c(T)\rightarrow0$ for decreasing temperature.
This behavior implies that the spectral function of the wide band exhibits
a pseudogap at $E_F$ which becomes progressively deeper towards low $T$.  

\begin{figure}[t!]
  \begin{center}
  \includegraphics[width=5.0cm,height=8cm,angle=-90]{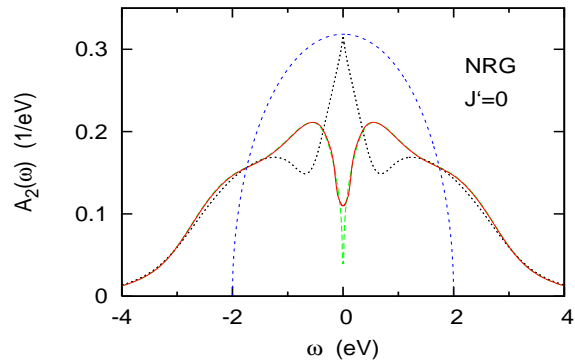}
  \end{center}
  \vskip-2mm
\caption{\label{NRG12J0}
Spectral distribution of wide band in intermediate phase, calculated 
within NRG approach for $U=2.4$~eV, $J=U/4$, $J'=0$.
Green curve: $T=3$~meV; red curve: $T=34$~meV;  
dashed blue curve: bare density of states. 
Black dotted curve: spectrum for $J'=J$, $T=3$~meV from 
Fig.~\ref{NRG12}. 
}\end{figure}

This picture is fully confirmed by the NRG results shown in 
Fig.~\ref{NRG12J0}. In contrast to the narrow peak at $E_F$ in the
three-peak structure seen in Fig.~\ref{NRG12} for isotropic Hund's 
coupling, the spectra for $J'=0$ show a pseudogap which becomes deeper 
as $T$ decreases. As a result, the spectral distribution now exhibits a 
characteristic four-peak structure, with two maxima limiting the pseudogap 
and two shoulders associated with the Hubbard peaks. Spectra of this kind 
were first observed in the QMC/DMFT results at $T=31$~meV reported in 
Ref.~\cite{prb70} (see also next section). The comparison with the
spectra for $J'=J$ reveals nearly identical excitations for 
$\vert\omega\vert\gtrsim J$. Thus, as expected, the different treatments 
of exchange interactions affect primarily the low-frequency excitations 
in the metallic wide band.

\section{7 \ Comparison with previous QMC/DMFT results}

In this and the following sections we compare the ED and NRG results 
for $J'=0$ with available QMC/DMFT data in order to illustrate 
the consistency between these impurity treatments and to explore 
further the role of finite-size effects in the ED approach.

\begin{figure}[t!]
  \begin{center}
  \includegraphics[width=5.9cm,height=8cm,angle=-90]{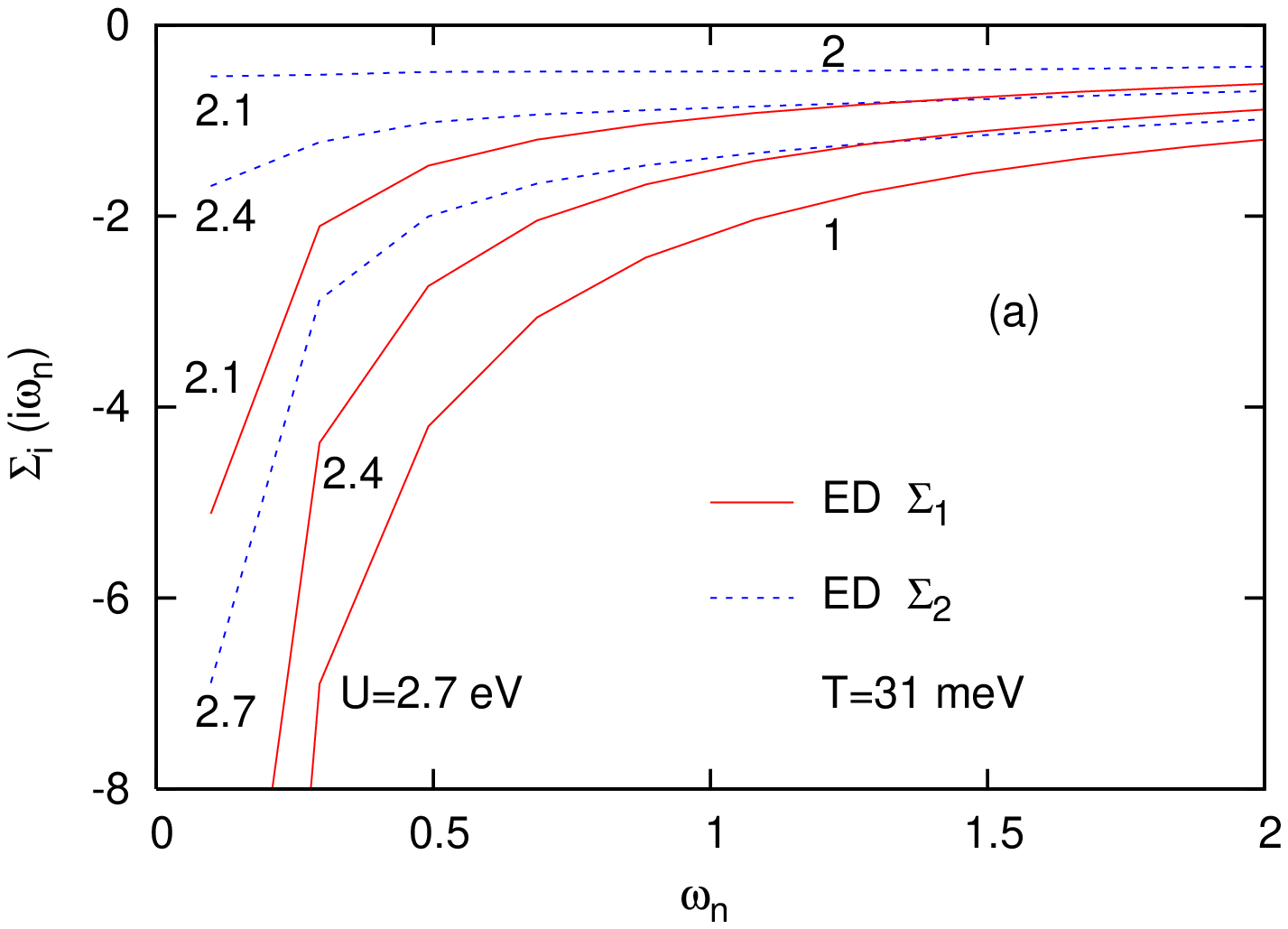}
  \includegraphics[width=5.9cm,height=8cm,angle=-90]{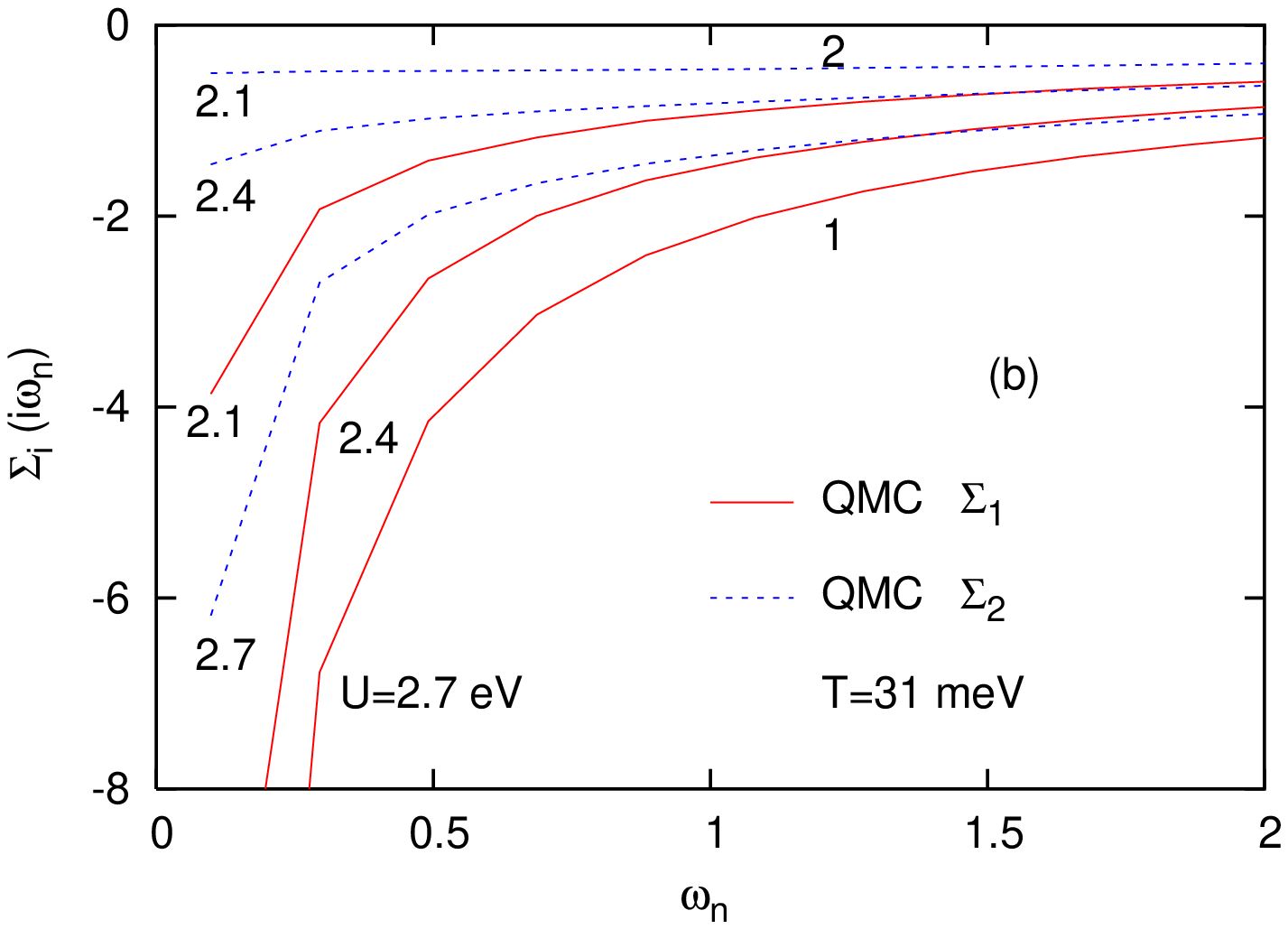}
  \includegraphics[width=5.9cm,height=8cm,angle=-90]{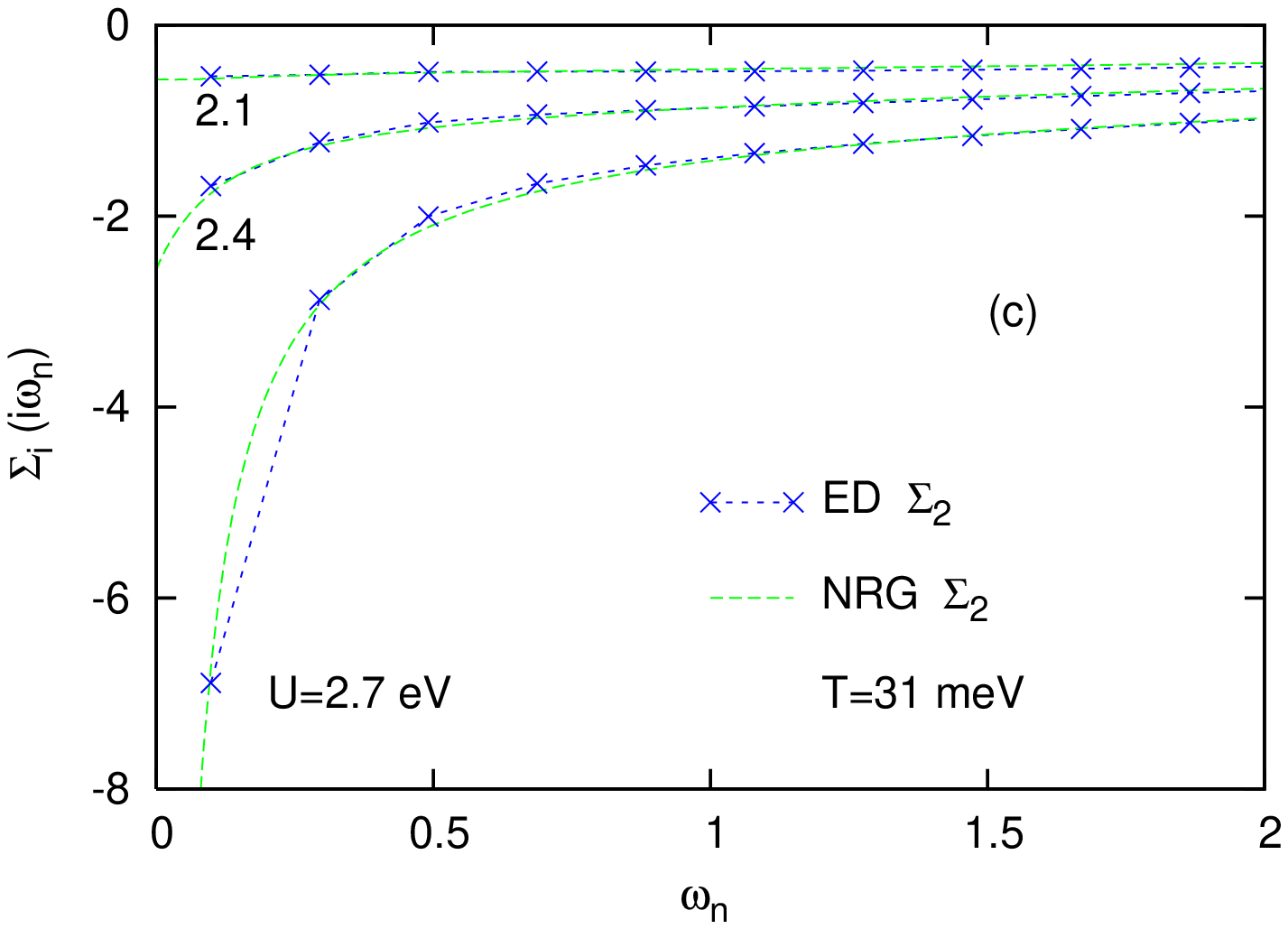}
  \end{center}
  \vskip-2mm
\caption{\label{QMCprb}
Comparison of subband self-energies $\Sigma_i(i\omega_n)$ in intermediate 
non-Fermi-liquid phase for $U=2.1,\ 2.4,\ 2.7$~eV, $J=U/4$, $J'=0$,
$T=31$~meV, calculated via three different impurity solvers:
(a) ED results for $n_s=8$: red solid curves: narrow band; 
blue dashed curves: wide band; 
(b) QMC results from Fig.~10 of Ref.~\cite{prb70};
red solid curves: narrow band; blue dashed curves: wide band;  
(c) comparison of ED (x) and NRG self-energies of wide band.
}\end{figure}

As shown in Ref.~\cite{prb70}, in the absence of spin-flip and
pair-exchange terms, purely metallic and insulating phases exist
for $U<U_{c1}\approx2.1$~eV and  $U>U_{c2}\approx2.7$~eV, 
respectively. Fig.~\ref{QMCprb} shows a comparison of subband 
self-energies at three representative Coulomb energies within 
the intermediate `bad-metal' non-Fermi-liquid region. The ED 
results for $n_s=8$ are seen to be in excellent agreement with 
the NRG self-energies, confirming the validity of the effective 
one-band model in the intermediate phase for $J'=0$. Moreover, 
both schemes agree very well with the QMC/DMFT self-energies 
reported in Ref.~\cite{prb70} for $T=31$~meV. Minor differences 
between the QMC and ED/NRG results are found only in the steepest 
parts of $\Sigma_i$ close to the first Matsubara frequency. 
(In contrast to the QMC calculations which are carried out at 
discrete Matsubara frequencies, the NRG self-energy is available 
continuously as a function of frequency. The ED self-energies 
could in principle also be obtained at arbitrary $i\omega$ but 
were calculated here at $i\omega_n$. This explains the slightly 
different form of the curves plotted in panel (c). At 
$i\omega_n$ the ED and NRG data nearly coincide.)  

\begin{figure}[t!]
  \begin{center}
  \includegraphics[width=5.0cm,height=8cm,angle=-90]{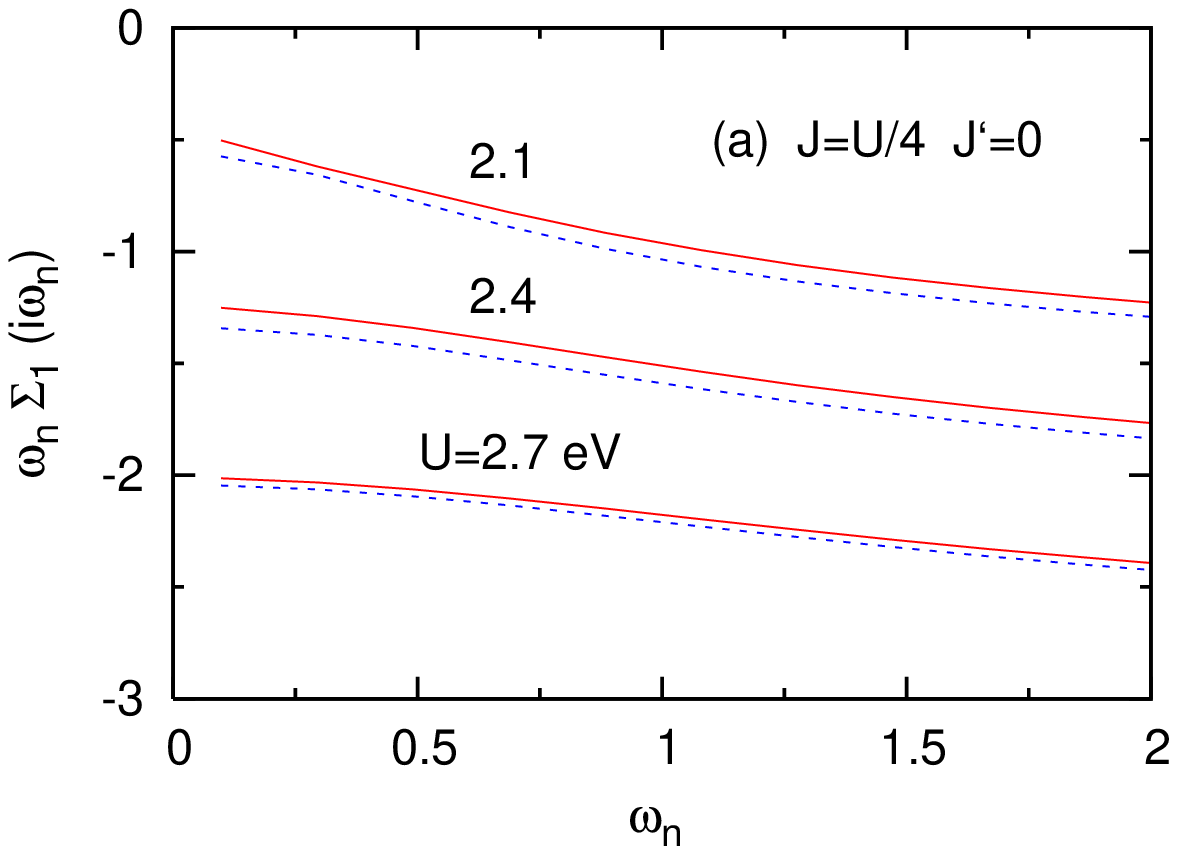}
  \includegraphics[width=5.0cm,height=8cm,angle=-90]{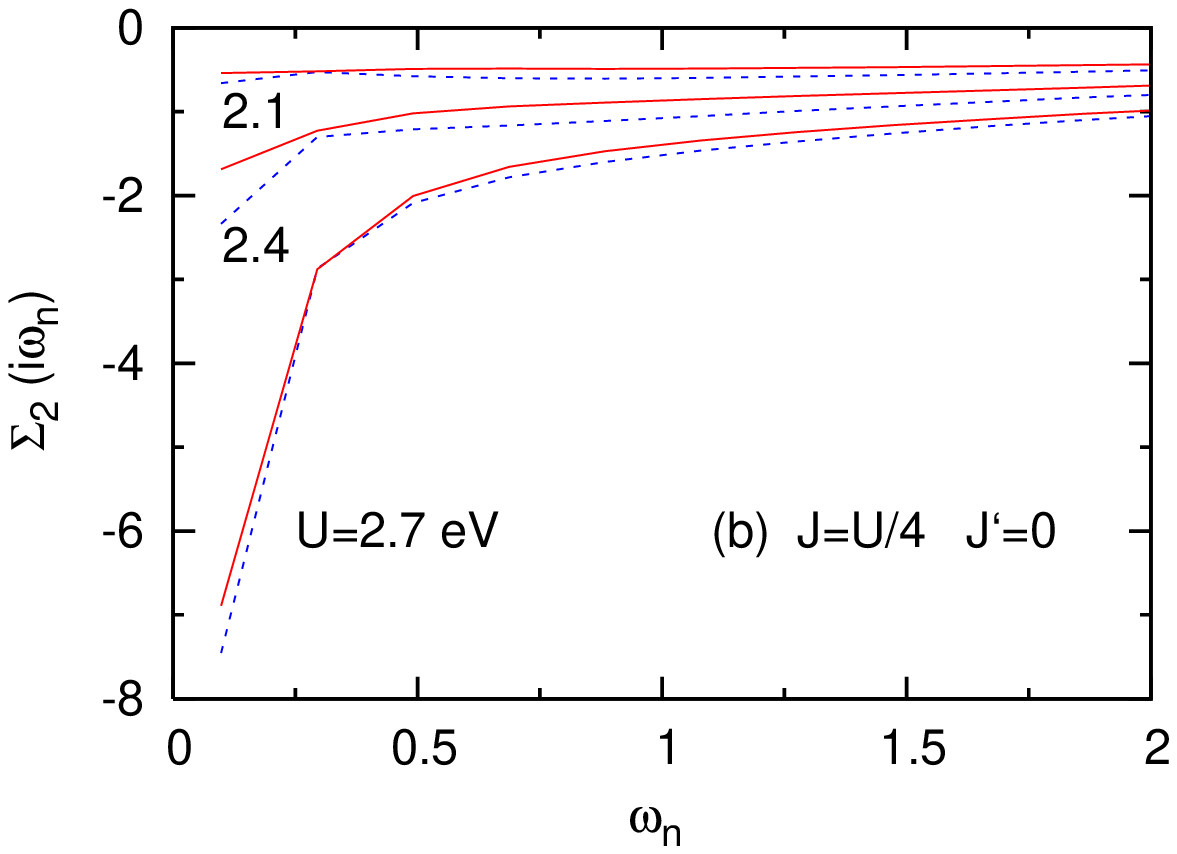}
  \end{center}
  \vskip-2mm
\caption{\label{QMCab}
Subband self-energies $\Sigma_i(i\omega_n)$ in intermediate 
non-Fermi-liquid phase for $T=31$~meV, $U=2.1,\ 2.4,\ 2.7$~eV,
$J=U/4$, $J'=0$. Solid (red) curves: $n_s=8$; dashed (blue) curves:   
$n_s=6$. 
(a) narrow subband: $\omega_n\Sigma_1(i\omega_n)$; 
(b) wide   subband: $\Sigma_2(i\omega_n)$.  
}\end{figure}

The overall 
trend obtained previously within the QMC/DMFT is fully confirmed 
by the new ED and NRG calculations: The narrow band is insulating 
throughout this range of Coulomb energies, whereas the wide band 
changes gradually from metallic to insulating via progressive 
non-Fermi-liquid behavior. Thus, for small $\omega_n$,   
Im\,$\Sigma_2(i\omega_n)\rightarrow c(U)$, 
where $c(U)\rightarrow-\infty$ at $U\approx2.7$~eV.   

Evidently, despite their intrinsic numerical uncertainties,
the three complementary impurity solvers: ED, NRG and QMC provide 
perfectly consistent descriptions of the electronic properties of 
the orbital-selective phase for Ising-like exchange. 

Since the temperature $T=31$~meV in these results is relatively 
high, finite-size effects tend to be less pronounced than in the cases 
discussed in the previous sections at lower $T$. This is illustrated 
in Fig.~\ref{QMCab} where we compare the above ED results for $n_s=8$ 
with those for $n_s=6$. The self-energies of the narrow band are 
nearly identical and satisfy 
$\omega_n\Sigma_1(i\omega_n)\rightarrow {\rm const.}$ at low 
frequencies. This is plausible since the additional zero-energy 
bath level for $n_s=8$ carries very little weight in the insulating 
state. Because of breakdown of Fermi-liquid behavior, the self-energy 
of the wide band approaches a finite value in the limit    
$\omega_n\rightarrow0$. The differences between the results for 
$n_s=6$ and $n_s=8$ are very small as long as this band is either 
nearly metallic, like at $U=2.1$~eV, or nearly insulating, like at 
$U=2.7$~eV. Slightly larger differences are found only in the middle 
of the bad-metallic region near $U=2.4$~eV. As discussed in the 
preceding section, the spectral function then has a more complicated
four-peak structure as a result of the narrow pseudogap at $E_F$,
which cannot be adequately represented using only two bath levels. 
This region could possibly be even more accurately described by using 
four bath levels for the wide band and two levels for the insulating 
narrow band, maintaining the total $n_s=8$.      

 \begin{figure}[t!]
  \begin{center}
  \includegraphics[width=6.1cm,height=8cm,angle=-90]{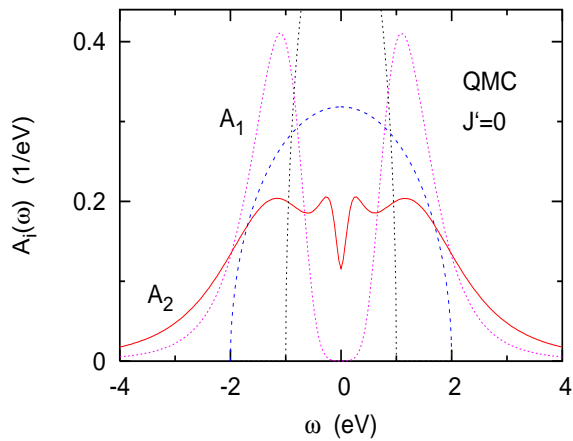}
  \end{center}
  \vskip-2mm
\caption{\label{QMCNRG}
Spectral distributions of subbands in intermediate phase, calculated 
within QMC/DMFT for $U=2.4$~eV, $J=U/4$, $J'=0$, $T=31$~meV.
Red curve: metallic wide band; magenta curve: insulating narrow band;
blue and black curves: bare densities of states. From Fig.~11(b) 
of Ref.~\cite{prb70}.
}\end{figure}
         
We close this section by showing in Fig.~\ref{QMCNRG} the spectral 
distributions of both subbands, as calculated within the QMC/DMFT and 
the maximum entropy method \cite{prb70}. The spectrum of the bad-metallic 
wide band can be compared with the corresponding NRG spectra plotted in 
Fig.~\ref{NRG12J0}. Both distributions exhibit the marked four-peak
structure induced by the pseudogap and the Hubbard bands. The 
low-frequency region is seen to be in excellent agreement. Both methods
coincide in that, at $T\approx30$~meV, the interacting density of 
states at $E_F$ of the wide band is only about one third of the 
noninteracting one. The slightly larger differences at higher energies,
in particular, the position and relative weight of the Hubbard bands, 
may be caused by the choice of fitting parameters in the maximum entropy 
procedure, and by the less accurate nature of the NRG approach at high 
energies. 

\section{8 \ Comparison with QMC/DMFT results for $\bf W_2=10W_1$}

The breakdown of Fermi-liquid behavior in the orbital-selective phase
with Ising-like exchange was recently also studied by Biermann 
{\it et al.}~\cite{biermann05} for the same two-band Hubbard model 
as above, except for $W_1=0.2$~eV and $W_2=2$~eV. 
The subbands were also assumed to be non-hybridizing and half-filled. 
QMC/DMFT calculations were carried out for $T=1/120\ldots1/40$~eV, 
$U=0.8$~eV, $J=0$ and $J=U/4$, with $J'=0$, i.e., in the absence of 
spin-flip and pair exchange terms. For $J=0$ as well as $J=U/4$, the 
self-energy of the narrow band was found to diverge at low frequencies, 
demonstrating that this band is insulating since $W_1\ll U$. The wide 
band is in the intermediate orbital-selective region between the purely 
metallic and insulating phases since $U\ll W_2$. The self-energy of this 
band showed a striking variation with the magnitude of $J$: 
For $J=0$, $\Sigma_2(i\omega_n)\sim i\omega_n$ at low frequencies, as 
expected for Fermi-liquid behavior. For  $J=U/4$, however, 
Im\,$\Sigma_2(i\omega_n)\rightarrow{\rm const.}\approx -0.09$ nearly 
independently of temperature for $T=1/120\ldots1/40$~eV, indicating 
breakdown of Fermi-liquid properties. Accordingly, the pinning condition 
$N_2(0)=A_2(0)$ was found to be satisfied for $J=0$, but not for $J=U/4$. 
These results are fully consistent with the trend discussed in 
Ref.~\cite{prb70} for $W_1=2$~eV and $W_2=4$~eV.   

\begin{figure}[t!]
  \begin{center}
  \includegraphics[width=5.0cm,height=8cm,angle=-90]{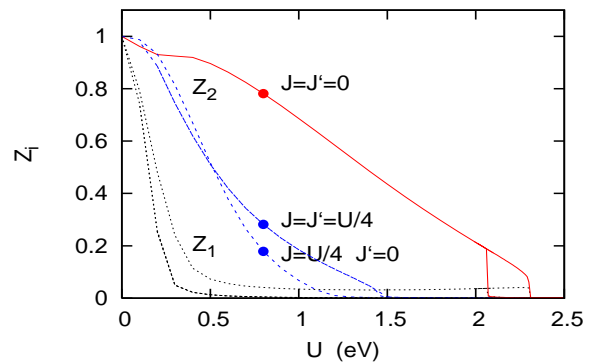}
  \end{center}
  \vskip-2mm
\caption{\label{Paris.zu}
$Z_i(U)$ as a function of $U$ for $W_1=0.2$~eV, $W_2=2.0$~eV, 
$T=1/120$~eV, calculated within ED/DMFT, $n_s=6$, for 
$J=J'=0$,  $J=J'=U/4$, and $J=U/4$, $J'=0$.    
Red and blue curves: wide band, black curves: narrow band.
The $Z_1$ for  $J=J'=U/4$ and $J=U/4$, $J'=0$ are indistinguishable. 
The dots mark the three cases shown in Fig.~\ref{Paris}.
}\end{figure}

To check the accuracy of our finite-$T$ ED approach we have applied 
it to the case investigated in Ref.~\cite{biermann05}. To provide a 
picture of the Mott transitions in this two-band system we show first 
in Fig.~\ref{Paris.zu} the variation of $Z_i(U)$ for three different 
treatments of Hund's exchange. In all cases the narrow band 
becomes insulating at about $U_{c1}=0.3\ldots0.4$~eV. 
However, the range and nature of the orbital-selective phase of the 
wide band, $U_{c1} < U < U_{c2}$, depend sensitively on the magnitudes 
of $J/U$ and $J'/U$. For $J=J'=0$ the upper transition occurs for 
$U_{c2}$ slightly larger than $W_2$, with a pronounced hysteresis 
loop indicative of first-order behavior. A very weak hysteresis 
is found also for $J=J'=U/4$, with $U_{c2}\approx 1.5$~eV, similar 
to the one in Fig.~\ref{ZU} near $U\approx3$~eV. Finally, for $J=U/4$ and 
$J'=0$, $U_{c2}\approx 1.2$~eV without evidence of first-order behavior.
(The latter result is consistent with $U_{c2}\approx 2.4$~eV in Fig.~2
of Ref.~\cite{prl05}.)  Although $Z_2(U)\approx 0.2\ldots0.8$ at 
$U=0.8$~eV, i.e., for $U_{c1}\ll U \ll U_{c2}$, the analysis of the 
self-energy reveals fundamentally different electronic properties 
of the wide band in this region.     

\begin{figure}[t!]
  \begin{center}
  \includegraphics[width=5.0cm,height=8cm,angle=-90]{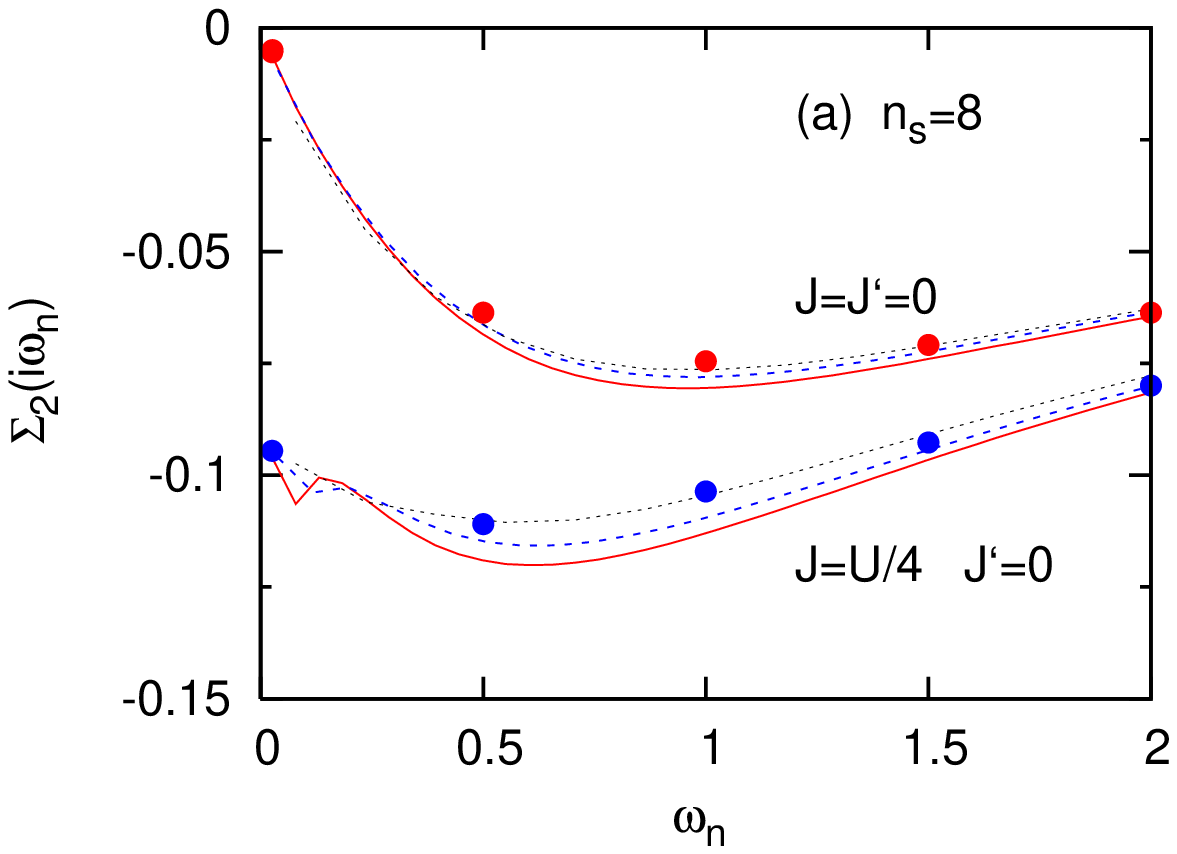}
  \includegraphics[width=5.0cm,height=8cm,angle=-90]{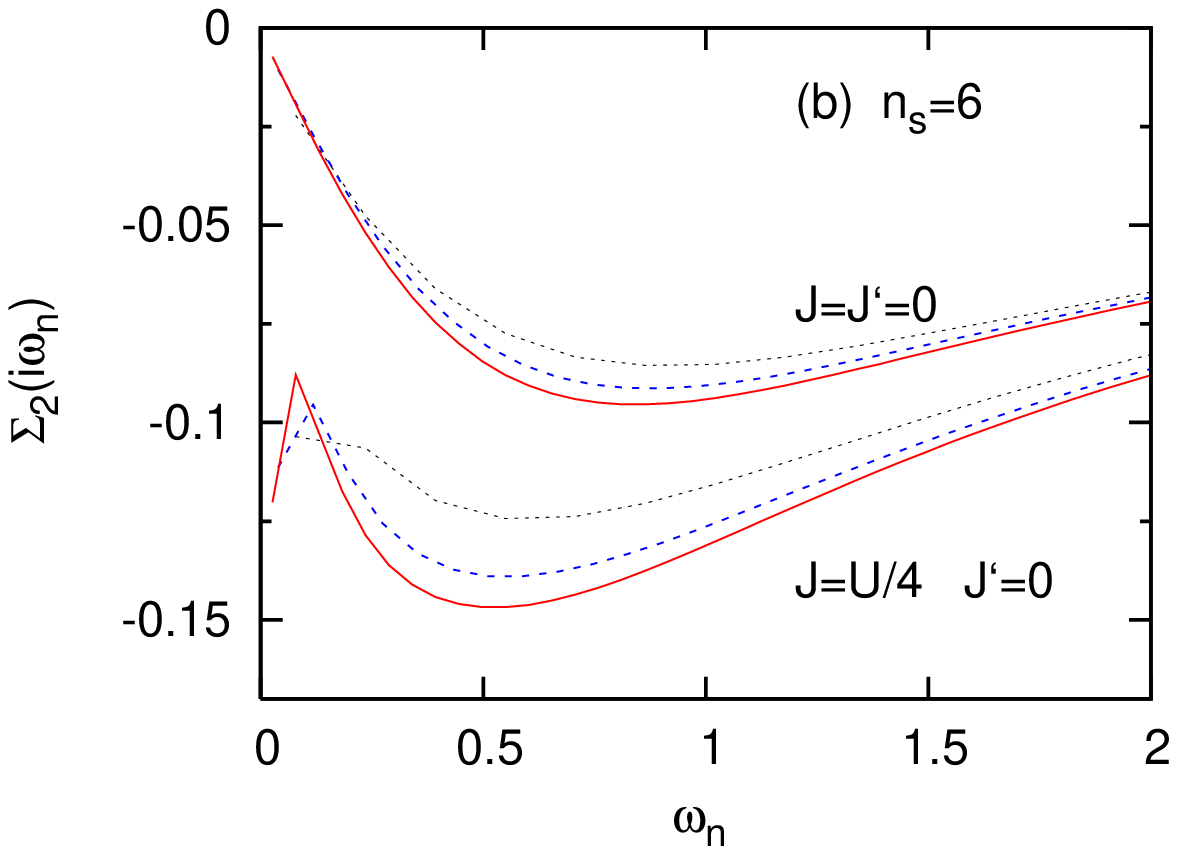}
  \includegraphics[width=5.0cm,height=8cm,angle=-90]{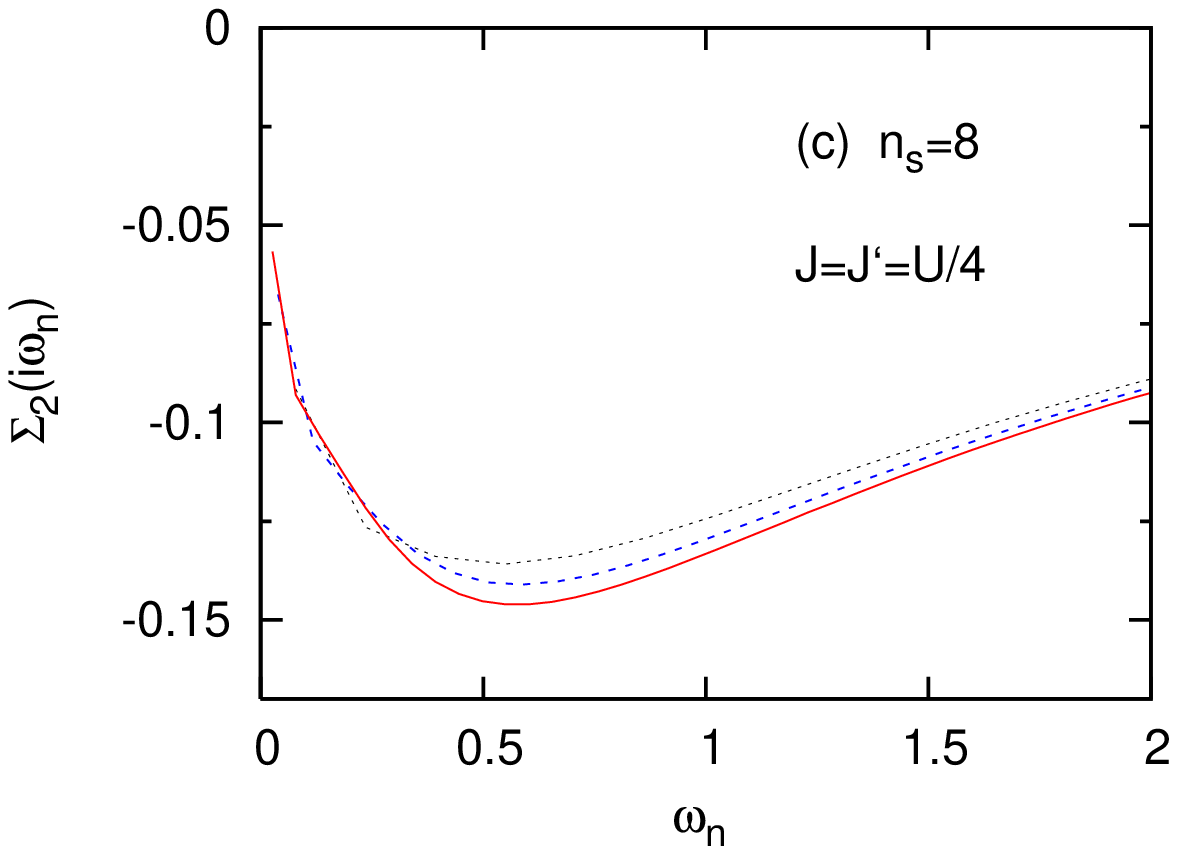}
  \end{center}
  \vskip-2mm
\caption{\label{Paris}
(a) Imaginary part of self-energy of wide band 
for $J=0$ and $J=U/4$ in intermediate 
phase for $U=0.8$~eV, $J'=0$, calculated within ED/DMFT with $n_s=8$. 
$W_1=0.2$~eV, $W_2=2.0$~eV.  Red solid, blue dashed, and black dotted 
curves: $T=1/120$~eV, $T=1/80$~eV, and $T=1/40$~eV, respectively.
The dots denote the results for $T=1/120$~eV obtained in 
Ref.~\cite{biermann05}. 
(b) Same as (a) except for $n_s=6$.  
(c) Same as (a) except for $J'=J=U/4$.
}\end{figure}

This is illustrated in Fig.~\ref{Paris} which shows 
Im\,$\Sigma_2(i\omega_n)$ for $n_s=8$ and $n_s=6$. The ED results for 
$n_s=8$ are in excellent agreement with the QMC data \cite{biermann05}
for $J=J'=0$ as well as $J=U/4$, $J'=0$, with the exception of small 
finite-size effects at the lowest Matsubara frequencies and lowest 
temperature. (Since $W_2=2$~eV, $T=1/120\ldots1/40$~eV in these 
calculations corresponds to about $T=16\ldots50$~meV in the case 
$W_2=4$~eV considered in the preceding sections.)  Also, the 
variation with $T$ is in our results slightly more pronounced 
than in the case of the QMC/DMFT.

For $n_s=6$ the ED results exhibit larger finite-size effects, as shown 
in Fig.~\ref{Paris}(b). The variation with temperature is in this case 
also larger than for $n_s=8$. Evidently, the subtle features of spectral 
functions in the intermediate phase are not so well represented by 
including only two bath levels. Nevertheless, the qualitative 
difference between Fermi-liquid behavior for $J=0$ and the clear 
deviation from this behavior for $J=U/4$ are very well reproduced 
by these $n_s=6$ ED/DMFT calculations.   

For completeness we show in Fig.~\ref{Paris}(c) the $n_s=8$ ED/DMFT 
results for isotropic Hund's coupling, i.e., $J'=J=U/4$. This case 
is not yet accessible within QMC/DMFT because of sign problems at low 
temperatures. 
The inclusion of spin-flip and pair-exchange terms is seen to restore
the limiting behavior $\Sigma_2(i\omega_n)\rightarrow 0$ for $\omega_n
\rightarrow 0$. Nevertheless, compared to the case $J'=J=0$ shown in (a),
$\Sigma_2$ now increases much more rapidly at small finite frequencies, 
similarly to the data shown in Figs.~\ref{OSMT} and \ref{NRG34}.  
(As argued in Section 5, the shoulder near $\omega_n\approx0.06$ is  
caused by finite-size effects.)
Thus, instead of quasiparticles with weight $Z_2\approx0.8$ for $J=J'=0$,
finite Hund's exchange with $J'=J=U/4$ supports a state of infinite 
lifetime at $E_F$, but does not satisfy Fermi-liquid criteria at 
$\omega>0$. This picture is consistent with the $T=0$ ED/DMFT results
by Biermann {\it et al.}~\cite{biermann05}.
The spectral function of the wide band therefore should exhibit 
a narrow peak at $E_F$, similar to the one shown in Fig.~\ref{NRG12} 
for $U=2.4$~eV.

\section{9 \ Comments on IPT/DMFT results}

In Ref.~\cite{prb70} we reported DMFT calculations for the two-band 
Hubbard model within QMC and iterated perturbation theory \cite{ipt}. 
Both approaches were shown to yield a consistent picture of the
Mott transition, in the sense that both exhibit a single first-order
transition at which the narrow band becomes insulating and the 
wide band begins to show progressive bad-metallic behavior.
This band becomes fully insulating at a higher Coulomb energy in a 
non-first-order manner. As shown via ED/DMFT at $T=0$ in Ref.~\cite{koga2} 
and at $T>0$ in Ref.~\cite{prl05}, the nature of this upper 
transition depends in a critical way on the treatment of onsite 
exchange interactions. In the absence of spin-flip and pair-exchange
terms it is continuous, consistent with the QMC results, while for
full Hund's coupling it becomes first-order.

The question then arises whether the IPT approach also supports this
picture. In fact, as pointed out in Ref.~\cite{prl05}, it is surprising 
that the IPT gives only one common first-order transition even though 
spin-flip and pair-exchange terms are included. It can easily be shown,
however, that to second-order in the Coulomb interaction the subband 
self-energies with and without spin-flip and pair-exchange are the same, 
except for a slightly different relation between $U$ and $J$. Thus, 
$J'=J=0.25\,U$ gives the same result as $J'=0$ and $J\approx0.22\,U$. 
Corrections to the simple second-order diagram via renormalization of subband
energies to yield the correct atomic limit \cite{kajueter,lichtenstein}
are also insensitive to the choice of $J'$ for the present two-band model. 

To resolve this puzzle we have repeated the IPT calculations reported 
in Ref.~\cite{prb70} at even lower temperatures and found indeed a tiny 
hysteresis loop also at the upper Mott transition, with a critical 
temperature of approximately $T_{c2}\approx5$~meV, i.e., significantly 
lower than the critical temperature of the lower Mott transition, 
$T_{c1}\approx50$~meV. These findings suggest that present formulations 
of IPT are too simple to deal with the full complexity of Hund's  
exchange. Diagrams beyond second-order are required to distinguish
more clearly between $J'=J$ and $J'=0$ treatments.

\section{10 \ Summary and outlook} 

Finite-temperature ED/DMFT studies are carried out in order to 
explore the usefulness of this approach for multi-band systems.
The important feature here is that onsite Coulomb and 
exchange interactions are fully included. Since in the past ED
has been used mainly for single-band systems, a particular aim 
of this work is to illustrate the dependence of self-energies 
on the cluster size and to test the range of applicability of 
this method. As an example, we focus on the Hubbard model for
two bands of different widths and investigate the metal insulator 
transition as a function of Hund's exchange.   

The surprising and potentially very useful result of this work is
that even a cluster size of $n_s=6$, with only two bath levels per 
impurity orbital, provides a qualitatively correct picture in all 
of the important phases of the $T/U$ phase diagram. Moreover, 
finite-size effects are substantially reduced for $n_s=8$, i.e., using
one extra bath level per impurity orbital. Thus, in the phase just 
below the main first-order Mott transition, both subbands exhibit 
clear metallic properties, albeit with strongly reduced 
quasi-particle weights. The intermediate region between the purely
metallic and insulating phases depends in a subtle manner on the 
exchange interactions included in the ED calculation.  
For full Hund's coupling coexisting metallic and insulating
subbands are found, where the wide band exhibits infinite lifetime
at $E_F$, but non-Fermi-liquid behavior at finite frequencies.
For Ising-like exchange, this breakdown of Fermi-liquid behavior
is enhanced, giving finite lifetime even at $E_F$. 

To explore the influence of finite-size effects in the two-band 
ED/DMFT approach NRG calculations were performed for an  effective 
one-band model suitable to describe the wide band in the intermediate 
phase. Both for full Hund's coupling and Ising-like exchange,
very good agreement with the ED results is found for 
temperatures in the range $T=20\ldots30$~meV. At lower $T$, larger
differences appear on a quantitative level at low frequencies.
Nevertheless, the important qualitative differences between various 
phases, especially the characteristic low-frequency variation of 
the self-energy in the two types of non-Fermi-liquid regions for 
isotropic and anisotropic exchange coupling, are fully reproduced 
by the ED approach, both for $n_s=6$ and $n_s=8$.     

To test the accuracy of the two-band ED approach we also have applied
it to models studied earlier within QMC/DMFT, neglecting spin-flip
and pair-exchange. Nearly quantitative agreement is obtained for
a cluster size $n_s=8$, but even the ED results for $n_s=6$ are
found to be qualitatively reliable. 

The present ED approach utilizes full diagonalization of the 
impurity Hamiltonian. Since at low temperatures only a limited 
range of excited states is relevant for the local Green's functions 
and self-energies it should be very useful to generalize finite-$T$ 
Lanczos one-band methods \cite{jaklic,capone} in order to bridge the 
gap between the present work and the true $T=0$ limit, and to apply 
the finite-temperature ED/DMFT approach to realistic two-band and 
three-band materials.

\bigskip
{\bf Acknowledgements}: \ 
One of us (A. L.) likes to thank A. I. Lichtenstein for parts of 
the multi-band ED code. Some of the ED and NRG DMFT calculations 
were carried out on the IBM supercomputer (JUMP) of the 
Forschungszentrum J\"ulich.

\bigskip
Email: \ a.liebsch@fz-juelich.de; \ t.costi@fz-juelich.de

\end{document}